\documentclass[preprint,showpacs,showkeys,superscriptaddress,floatfix,
prd,amsmath,amssymb]{revtex4}

\usepackage{graphicx}% Include figure files
\usepackage{dcolumn}% Align table columns on decimal point
\usepackage{bm}% bold math

\newcommand{\onehalf}{{\textstyle \frac{1}{2}}}
\newcommand{\third}{{\textstyle \frac{1}{3}}}
\newcommand{\twothird}{{\textstyle \frac{2}{3}}}
\newcommand\cO{\mathcal{O}} 

\begin{document}

\title{Nucleon electromagnetic form factors on the lattice 
       and in chiral effective field theory}

\author{M. G\"ockeler}
\affiliation{Institut f\"ur Theoretische Physik,
             Universit\"at Leipzig, D-04109 Leipzig, Germany}
\affiliation{Institut f\"ur Theoretische Physik,
             Universit\"at Regensburg, D-93040 Regensburg, Germany}
\author{T.R. Hemmert}
\affiliation{Physik-Department, Theoretische Physik,
             Technische Universit\"at M\"unchen, D-85747 Garching, Germany}
\author{R. Horsley}
\affiliation{School of Physics, University of Edinburgh,
             Edinburgh EH9 3JZ, UK}
\author{D. Pleiter}
\affiliation{John von Neumann-Institut f\"ur Computing NIC,
             Deutsches Elektronen-Synchrotron DESY, D-15738 Zeuthen, Germany}
\author{P.E.L. Rakow}
\affiliation{Theoretical Physics Division, 
             Department of Mathematical Sciences,
             University of Liverpool, Liverpool L69 3BX, UK}
\author{A. Sch\"afer}
\affiliation{Institut f\"ur Theoretische Physik,
             Universit\"at Regensburg, D-93040 Regensburg, Germany}
\author{G. Schierholz}
\affiliation{John von Neumann-Institut f\"ur Computing NIC,
             Deutsches Elektronen-Synchrotron DESY, D-15738 Zeuthen, Germany}
\affiliation{Deutsches Elektronen-Synchrotron DESY, D-22603 Hamburg, Germany}
\collaboration{QCDSF collaboration} \noaffiliation

\date{\today}% It is always \today, today,
             %  but any date may be explicitly specified

\begin{abstract}
We compute the electromagnetic form factors of the nucleon in quenched
lattice QCD, using non-perturbatively improved Wilson fermions,
and compare the results with phenomenology and chiral effective field theory.
\end{abstract}



\maketitle

\section{Introduction}
\label{sect:intro}
 
The measurements of the electromagnetic form factors of the 
proton and the neutron gave the first hints at an internal structure 
of the nucleon, and a theory of the nucleon cannot be considered
satisfactory if it is not able to reproduce the form factor data.
For a long time, the overall trend of the experimental results 
for small and moderate values of the momentum transfer $q^2$ 
could be described reasonably well by phenomenological (dipole) fits
\begin{eqnarray} 
G_e^p(q^2) &\sim& \frac{G_m^p(q^2)}{\mu^p} \sim \frac{G_m^n(q^2)}{\mu^n} 
\nonumber \\
{} &\sim& \left( 1 - q^2 / m_D^2 \right)^{-2} \,, \nonumber \\
G_e^n(q^2) &\sim& 0 \label{dipole}
\end{eqnarray}
with $m_D \sim 0.84 \, \text{GeV}$ and the magnetic moments
\begin{equation}
 \mu^p \sim 2.79  \quad , \quad
 \mu^n \sim -1.91 
\end{equation}
in units of nuclear magnetons.
Recently, the form factors of the nucleon have been studied experimentally
with high precision and deviations from this uniform dipole form have been
observed, both at very small $q^2$~\cite{gao} and in the region
above $1 \, \text{GeV}^2$~\cite{JLAB1,JLAB2}.

It is therefore of great interest to derive the nucleon form factors
from QCD. Since form factors are typical low-energy quantities, 
perturbation theory in terms of quarks and gluons
is useless for this purpose and a non-perturbative
method is needed. If one wants to avoid additional assumptions or models,
one is essentially restricted to lattice QCD and Monte Carlo simulations.
In view of the importance of nucleon form factors and the amount of 
experimental data available, it is surprising that there are only
a few lattice investigations of form factors in the last years~\cite{liu,ji}. 

In this paper we give a detailed account of our results for the nucleon 
form factors obtained in quenched Monte Carlo simulations with 
non-perturbatively $O(a)$-improved Wilson fermions. We shall pay particular
attention to the chiral extrapolation and compare with 
formulae from chiral effective field theory previously used for studies 
of the magnetic moments. The plan of the paper is 
as follows. After a few general definitions and remarks in the next section
we describe the lattice technology that we used in Sec.~\ref{sect:lattice}.
After commenting on our earlier attempts to perform a chiral extrapolation in
Sec.~\ref{sect:first}, we investigate the quark-mass dependence of the
form factors in more detail on a phenomenological basis in 
Sec.~\ref{sect:invest}. We find that our data are in qualitative agreement 
with the recently observed deviations~\cite{JLAB1,JLAB2} from the 
uniform dipole parametrization of the proton form factors.
In Sec.~\ref{sect:chpt} we collect and discuss formulae from chiral
effective field theory, which are confronted with our Monte Carlo
results in Sec.~\ref{sect:disc}. Our conclusions are presented 
in Sec.~\ref{sect:concl}. The Appendixes contain a short discussion of 
the pion mass dependence of the nucleon mass as well as tables of the 
form factors and of the corresponding dipole fits.

\section{Generalities}
\label{sect:general}

Experimentally, the nucleon form factors are measured via electron 
scattering. 
Because the fine structure constant is so small, it is justified to describe
this process in terms of one-photon exchange. So the 
scattering amplitude is given by
\begin{equation}
T_{fi} = e^2 \bar{u}_e (k'_e,s'_e) \gamma_\mu 
            u_e (k_e,s_e) \frac{1}{q^2}
\langle p',s' | J^\mu | p,s \rangle
\end{equation}
with the electromagnetic current 
\begin{equation}
J^\mu = \frac{2}{3} \bar{u} \gamma^\mu u - \frac{1}{3} \bar{d} \gamma^\mu d
 + \cdots \,.
\end{equation}
Here $p$, $p'$ are the nucleon momenta, $k_e$, $k'_e$ are the electron
momenta, and $s$, $s'$, \ldots are the corresponding spin vectors. The
momentum transfer is defined as $q=p'-p$. With the help of the form
factors $F_1 (q^2)$ and $F_2 (q^2)$ the nucleon matrix element can 
be decomposed as 
\begin{equation} \label{ffdef} 
 \langle p',s' | J^\mu | p,s \rangle 
 = \bar{u} (p',s') 
\left [\gamma_\mu F_1 (q^2) + 
   \mathrm i \sigma^{\mu \nu} \frac{q_\nu}{2M_N} F_2 (q^2) \right] u(p,s)
 \end{equation} 
where $M_N$ is the mass of the nucleon. From the kinematics of the scattering
process, it can easily be seen that $q^2 < 0$. In the following, we shall
often use the new variable $Q^2 = - q^2$. We have $F_1(0) = 1$ 
(in the proton) as $J$ 
is a conserved current, while $F_2(0)$ measures the anomalous magnetic 
moment in nuclear magnetons. For a classical point particle, both 
form factors are independent of $q^2$, so deviations from this behavior 
tell us something
about the extended nature of the nucleon. 
In electron scattering, $F_1$ and $F_2$ are usually re-written 
in terms of the electric and magnetic Sachs form factors
\begin{eqnarray}
G_e(q^2) &=& F_1(q^2) + \frac{q^2}{(2M_N)^2} F_2(q^2)\,, \nonumber \\
G_m(q^2) &=& F_1(q^2) + F_2(q^2) \,, \label{sachs} 
\end{eqnarray}
as then the (unpolarized) cross section becomes a linear combination
of squares of the form factors.

Throughout the whole paper we assume flavor SU(2) symmetry. Hence
we can decompose the form factors into isovector and isoscalar components.
In terms of the proton and neutron form factors the isovector
form factors are given by
\begin{eqnarray} 
G_e^v(q^2) &=& G_e^p(q^2) - G_e^n(q^2) \,, \nonumber \\
G_m^v(q^2) &=& G_m^p(q^2) - G_m^n(q^2) 
\end{eqnarray}
such that
\begin{equation} 
G_m^v(0) = G_m^p(0) - G_m^n(0) = \mu^p - \mu^n = \mu^v = 1 + \kappa_v 
\end{equation}
with $\mu^v$ ($\kappa_v$) being the isovector (anomalous) magnetic moment
$\sim 4.71 \: (3.71)$.
In the actual simulations we do not work directly with these definitions
when we calculate the isovector form factors. 
We use the relation
\begin{multline} \label{isovector}
 \langle \text{proton}| \left( \twothird \bar{u} \gamma^\mu u \right.
     -  \left. \third \bar{d} \gamma^\mu d \right) | \text{proton} \rangle
  -  \langle \text{neutron}| \left( \twothird \bar{u} \gamma^\mu u 
    - \third \bar{d} \gamma^\mu d \right) | \text{neutron} \rangle 
 \\
  {} =   \langle \text{proton}| \left( \bar{u} \gamma^\mu u 
    - \bar{d} \gamma^\mu d \right) | \text{proton} \rangle
\end{multline}
and compute the isovector form factors from proton matrix elements
of the current $\bar{u} \gamma^\mu u - \bar{d} \gamma^\mu d$ instead of
evaluating the proton and neutron matrix elements of the electromagnetic
current separately and then taking the difference.
Similarly one could use the isoscalar current 
$\bar{u} \gamma^\mu u + \bar{d} \gamma^\mu d$ for the computation
of isoscalar form factors, but we have not done so, since in the isoscalar
sector there are considerable uncertainties anyway due to the neglected 
quark-line disconnected contributions.

\section{Lattice technology}
\label{sect:lattice}

\begin{table*}
\caption{Simulation parameters, numbers of gauge field configurations
used (\#~configs.) and masses.}
\label{tab:param}
\begin{ruledtabular}
\begin{tabular}{llllllll}
\multicolumn{1}{c}{$\beta $} & \multicolumn{1}{c}{$c_{SW}$} & 
\multicolumn{1}{c}{$\kappa $} & \multicolumn{1}{c}{$c_V$} & 
\multicolumn{1}{c}{Volume} & \multicolumn{1}{c}{\# configs.} &
\multicolumn{1}{c}{$a m_\pi $} & \multicolumn{1}{c}{$a M_N $} \\
\hline
6.0 & 1.769 & 0.1320 & $-0.331$ & $16^3 \times 32$ & $O(450)$ & 0.5412(9) 
& 0.9735(40) \\
6.0 & 1.769 & 0.1324 & $-0.331$ & $16^3 \times 32$ & $O(550)$ & 0.5042(7) 
& 0.9353(25) \\
6.0 & 1.769 & 0.1333 & $-0.331$ & $16^3 \times 32$ & $O(550)$ & 0.4122(9) 
& 0.8241(34) \\
6.0 & 1.769 & 0.1338 & $-0.331$ & $16^3 \times 32$ & $O(500)$ & 0.3549(12)
& 0.7400(85) \\ 
6.0 & 1.769 & 0.1342 & $-0.331$ & $16^3 \times 32$ & $O(700)$ & 0.3012(10)
& 0.7096(48) \\
6.0 & 1.769 & $\kappa_c = 0.1353$ & {} & {} & {} & {} & 0.5119(67) \\
\hline
6.2 & 1.614 & 0.1333 & $-0.169$ & $24^3 \times 48$ & $O(300)$ & 0.4136(6) 
& 0.7374(21) \\
6.2 & 1.614 & 0.1339 & $-0.169$ & $24^3 \times 48$ & $O(300)$ & 0.3565(8) 
& 0.6655(28) \\
6.2 & 1.614 & 0.1344 & $-0.169$ & $24^3 \times 48$ & $O(300)$ & 0.3034(6) 
& 0.5963(29) \\
6.2 & 1.614 & 0.1349 & $-0.169$ & $24^3 \times 48$ & $O(500)$ & 0.2431(7) 
& 0.5241(39) \\ 
6.2 & 1.614 & $\kappa_c = 0.1359$ & {} & {} & {} & {} & 0.3695(36) \\
\hline
6.4 & 1.526 & 0.1338 & $-0.115$ & $32^3 \times 48$ & $O(200)$ & 0.3213(8) 
& 0.5718(28) \\
6.4 & 1.526 & 0.1342 & $-0.115$ & $32^3 \times 48$ & $O(100)$ & 0.2836(9) 
& 0.5266(31) \\
6.4 & 1.526 & 0.1346 & $-0.115$ & $32^3 \times 48$ & $O(200)$ & 0.2402(8) 
& 0.4680(37) \\
6.4 & 1.526 & 0.1350 & $-0.115$ & $32^3 \times 48$ & $O(300)$ & 0.1933(7) 
& 0.4156(34) \\ 
6.4 & 1.526 & 0.1353 & $-0.115$ & $32^3 \times 64$ & $O(300)$ & 0.1507(8) 
& 0.3580(47) \\
6.4 & 1.526 & $\kappa_c = 0.1358$ & {} & {} & {} & {} & 0.2800(53)   
\end{tabular}
\end{ruledtabular}
\end{table*}

\begin{figure}
  \includegraphics[width=12.0cm]{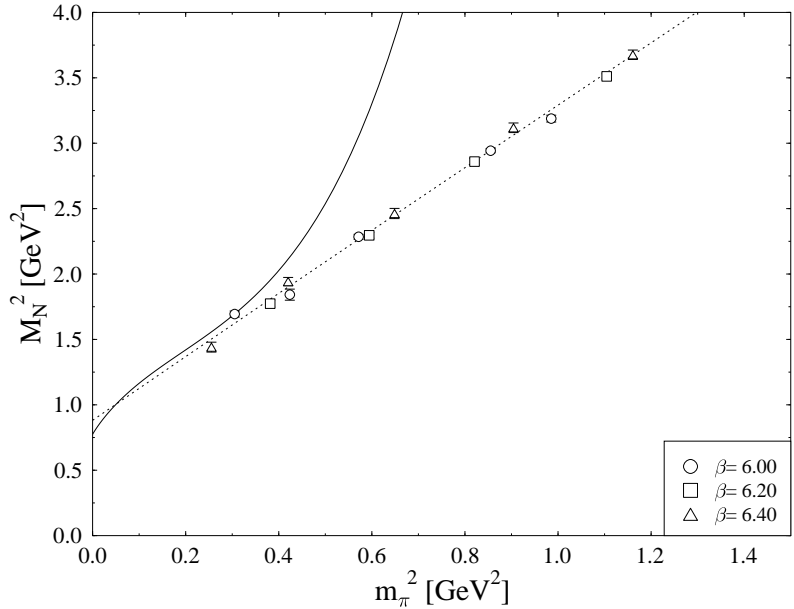}
    \caption{Nucleon mass squared versus pion mass squared from
             the data in Table~\ref{tab:param}. The dotted curve
             shows our phenomenological chiral extrapolation 
             (Eq.~(\ref{extra})) for $\beta=6.4$. The full curve 
             corresponds to the chiral extrapolation derived from 
             chiral perturbation theory (Eq.~(\ref{nuclmass})) with 
             the parameters given in Appendix~A.}
    \label{fig:masses}
\end{figure}

Using the standard Wilson gauge field action we have performed 
quenched simulations with 
$O(a)$-improved Wilson fermions (clover fermions). For
the coefficient $c_{SW}$ of the Sheikholeslami-Wohlert clover term we chose a
non-perturbatively determined value 
calculated from the interpolation formula given in Ref.~\cite{alpha}.
The couplings $\beta = 6/g^2$ and $c_{SW}$,
the lattice sizes and statistics, the values of the hopping parameter 
$\kappa$ (not to be confused with an anomalous magnetic moment) and the  
corresponding pion and nucleon masses (in lattice units) are collected 
in Table~\ref{tab:param}. 
As we are going to investigate nucleon properties, we want to determine
the lattice spacing from the (chirally extrapolated) nucleon mass in order
to ensure that the nucleon mass takes the correct value. (At the present 
level of accuracy, the difference between the nucleon mass in the chiral 
limit and the physical nucleon mass can be neglected.) Ideally we would 
use a formula from chiral perturbation theory for this purpose, e.g.\
Eq.~(\ref{nuclmass}). Since there seems to be little
difference between quenched and unquenched results at presently accessible 
quark masses it would make sense to apply this formula to our 
data. However, it turns out that it breaks down for pion masses above 
600 MeV~\cite{bhm}, where almost all of our results lie 
(see Fig.~\ref{fig:masses}). 
(For a detailed discussion of a different approach see Ref.~\cite{lein}.)
Hence we resort to a simple-minded phenomenological procedure
extrapolating our masses by means of the ansatz
\begin{equation} \label{extra}
(a M_N)^2 = (a M_N^0)^2 + b_2 (a m_\pi)^2 + b_3 (a m_\pi)^3 
\end{equation}
for each $\beta$. This ansatz provides a very good description of the
data~\cite{dirk}. The nucleon masses extrapolated to the critical hopping 
parameter $\kappa_c$ in this way (on the basis of a larger set of nucleon 
masses) are also given in Table~\ref{tab:param}. Whenever we give numbers 
in physical units the scale has been set by these chirally extrapolated 
nucleon masses.

In the last years it has become more popular to set the scale with 
the help of the force parameter $r_0$~\cite{sommer}. While this choice 
avoids the problems related to the chiral extrapolation of the nucleon 
mass, it has the disadvantage that the physical value of $r_0$ is less 
precisely determined than the physical nucleon mass. Furthermore, as 
the present paper deals exclusively with nucleon properties it seemed 
to us more important to have the correct value in physical units for 
the mass of the particle studied when evaluating other dimensionful 
quantities like, e.g., radii. It is however interesting to note that 
the dimensionless product $r_0 M_N^0$ is to a good accuracy independent 
of the lattice spacing. Indeed, taking $r_0$ from Ref.~\cite{guagnelli} and 
multiplying by the chirally extrapolated nucleon
masses given in Table~\ref{tab:param} one finds $r_0 M_N^0 = 2.75$, $2.72$, 
and $2.73$ for $\beta = 6.0$, $6.2$, and $6.4$, respectively. Thus the
scaling behavior of our results is practically the same for both choices
of the scale.

In Fig.~\ref{fig:masses} we plot $M_N^2$ versus $m_\pi^2$ using 
the masses from Table~\ref{tab:param}. The picture demonstrates that 
scaling violations in the masses are small. Moreover, we see that for
$m_\pi < 600 \, \text{MeV}$ our extrapolation curve is quite close to the
chiral perturbation theory curve (\ref{nuclmass}).

In order to compute nucleon masses or matrix elements we need suitable 
interpolating fields. For a proton with spatial momentum $\vec{p}$
the most obvious choice in terms of the quark fields $u(x)$ and $d(x)$ is
\begin{eqnarray} 
  B_\alpha (t,\vec{p}) & = & 
\sum_{x, \; x_4 = t}
\text{e}^{- {\mathrm i} \vec{p}\cdot \vec{x} }
\epsilon_{i j k} u^i_\alpha (x) u^j_\beta (x) 
(C \gamma_5)_{\beta \gamma} d^k_\gamma (x) \,,
\nonumber \\ 
  \bar{B}_\alpha (t,\vec{p}) & = &
\sum_{x, \; x_4 = t }
\text{e}^{{\mathrm i}\vec{p}\cdot \vec{x} }
\epsilon_{i j k} \bar{d}^i_\beta (x) (C \gamma_5)_{\beta \gamma} 
\bar{u}^j_\gamma (x) \bar{u}^k_\alpha (x) 
\label{nucfield} 
\end{eqnarray} 
with the charge conjugation matrix $C$ ($\alpha$, $\beta$, $\gamma$ are 
Dirac indices, $i$, $j$, $k$ are color indices). Note that we now
switch from Minkowski space to Euclidean space.

In Eq.~(\ref{nucfield}) all three quarks sit at the
same point. Clearly, as protons are not point objects this is not the best
thing to do, and with the above interpolating fields we run the risk that 
the amplitudes of one-proton states in correlation functions might be 
very small making the extraction of masses and matrix elements rather
unreliable. Therefore we employ two types of improvement: First we smear
the sources and the sinks for the quarks in their time slices, secondly
we apply a ``non-relativistic'' projection. 

Our smearing algorithm
(Jacobi smearing) is described in Ref.~\cite{pionrho}. The parameters
$N_{\mathrm {smear}}$, $\kappa_{\mathrm {smear}}$ used in the actual 
computations and the resulting 
smearing radii are given in Table~\ref{tab:smear}. A typical rms nucleon
radius is about 0.8 fm, our smearing radii are about half that size.
\begin{table}
\caption{Smearing parameters for Jacobi smearing.}
\label{tab:smear}
\begin{ruledtabular}
\begin{tabular}{cccc}
$\beta$   & $N_{\mathrm {smear}}$  & $\kappa_{\mathrm {smear}}$ 
& $r_{\mathrm{rms}}$  \\
\hline 
6.0       & 50     & 0.21        & $\sim 3.5 a \sim 0.38\,$fm \\
6.2       & 100    & 0.21        & $\sim 5.6 a \sim 0.44\,$fm \\
6.4       & 150    & 0.21        & $\sim 6.7 a \sim 0.40\,$fm 
\end{tabular}
\end{ruledtabular}
\end{table}

The ``non-relativistic'' projection means that we replace each spinor by
\begin{equation} 
\psi \to \psi^{NR} = \onehalf (1+\gamma_4) \psi \,, \,
\bar{\psi} \to \bar{\psi}^{NR} = \bar{\psi} \onehalf (1+\gamma_4) \,.
\end{equation}
This replacement leaves quantum numbers unchanged, but we would expect it
to improve the overlap with baryons. 
Practically this means that for each baryon propagator we 
consider only the first two Dirac components. So we 
only have $2 \times 3$ inversions to perform rather than the usual
$4 \times 3$ inversions -- a saving of 50\% in computer time. 

The non-forward matrix elements required for the form factors are computed
from ratios of three-point functions to two-point functions. The two-point
function is defined as 
\begin{equation}
C_2(t,\vec{p}\,) = \sum_{\alpha \beta} \Gamma_{\beta \alpha}
\langle B_\alpha (t,\vec{p}\,) \bar{B}_\beta (0,\vec{p}\,) \rangle
\end{equation} 
with the spin projection matrix
\begin{equation}
\Gamma = \onehalf ( 1 + \gamma_4) \,.
\label{Gunpol}
\end{equation} 
In the three-point function 
\begin{equation}
C_3(t,\tau,\vec{p},\vec{p}\,') = \sum_{\alpha \beta} \Gamma_{\beta \alpha}
\langle B_\alpha (t,\vec{p}\,) \cO (\tau) \bar{B}_\beta (0,\vec{p}\,') \rangle
\end{equation} 
we have used, besides the matrix (\ref{Gunpol}) corresponding to unpolarized
matrix elements, also 
\begin{equation} \label{Gpol}
\Gamma = \onehalf ( 1 + \gamma_4) \mathrm i \gamma_5 \gamma_2
\end{equation} 
corresponding to polarization in the 2-direction. We then computed the
ratios 
\begin{equation} \label{ratio} 
 R(t,\tau,\vec{p},\vec{p}\,') =  
\frac{C_3(t,\tau,\vec{p},\vec{p}\,')}{C_2(t,\vec{p}\,)} \times
 \left[ \frac{C_2(\tau,\vec{p}\,) C_2(t,\vec{p}\,) C_2(t-\tau,\vec{p}\,') }
{C_2(\tau,\vec{p}\,') C_2(t,\vec{p}\,') C_2(t-\tau,\vec{p}\,)} 
\right]^{1/2} \,.
\end{equation} 
If all time differences are sufficiently large, i.e.\ if $0 \ll \tau \ll t$,
$R$ is proportional to the (polarized or unpolarized) proton 
matrix element of the operator $\cO$ with a known kinematical coefficient
presented below.

For the electromagnetic form factors the operator to be studied is the
vector current. In contrast to previous investigations~\cite{liu,ji}, 
which used the conserved vector current, we chose to work with the local 
vector current $\bar{\psi}(x) \gamma_\mu \psi(x)$.
The local vector current has to be renormalized, because it is not conserved.
It should also be improved so that its matrix elements 
have discretization errors of $O(a^2)$ only, which means that we use 
the operator
\begin{equation} 
 V_\mu = Z_V (1+b_V a m_q) \left[ \bar{\psi} \gamma_\mu \psi 
   + {\mathrm i} c_V a \partial_\lambda 
      ( \bar{\psi} \sigma_{\mu \lambda} \psi ) \right] \,,
\end{equation}
where $m_q$ is the bare quark mass:
\begin{equation} 
a m_q = \frac{1}{2 \kappa} - \frac{1}{2 \kappa_c} \,.
\end{equation}
We have taken $Z_V$ and $b_V$ from the parametrizations
given by the ALPHA collaboration~\cite{zvclover} (see also 
Ref.~\cite{zvroger}). The improvement coefficient $c_V$ has also
been computed non-perturbatively~\cite{cv}. The results can be
represented by the expression~\cite{dirk}
\begin{equation} 
c_V = - 0.01225 \,\frac{4}{3} \, g^2 \frac{1 - 0.3113 g^2}{1 - 0.9660 g^2} \,,
\end{equation}
from which we have calculated $c_V$ (see Table~\ref{tab:param}). 
In the limit $g^2 \to 0$ it agrees
with perturbation theory~\cite{weisz}. Computing all these additional
contributions in our simulations, we found the improvement terms to be
numerically small.
Note that the improvement coefficient $c_{CVC}$ for the conserved vector
current is only known to tree level so that a fully non-perturbative 
analysis would not be possible had we used the conserved vector current.

In order to describe the relation between the ratios we computed and the
form factors let us call the ratio $R$ for the $\mu$-component of the 
renormalized vector current more precisely $R_\mu$. Furthermore we
distinguish the unpolarized case (spin projection matrix (\ref{Gunpol})) 
from the polarized case (spin projection matrix (\ref{Gpol})) by a 
superscript. The (Minkowski) momentum transfer is given by
\begin{equation} 
q^2 = - Q^2 = 2 \left ( M_N^2 + \vec{p} \cdot \vec{p}\,' - 
     E_N(\vec{p}\,) E_N(\vec{p}\,') \right )
\end{equation}
with the nucleon energy
\begin{equation} 
E_N(\vec{p}\,) = \sqrt{M_N^2 + \vec{p}\,^2} \,.
\end{equation}
Using the abbreviation
\begin{equation} 
A(\vec{p},\vec{p}\,')^{-1} = (- Q^2 - 4 M_N^2) 
\sqrt{E_N(\vec{p}\,) (M_N + E_N(\vec{p}\,))
  E_N(\vec{p}\,') (M_N + E_N(\vec{p}\,'))} 
\end{equation}
we have
\begin{multline} \label{r4unpol}
R_4^{\mathrm {unpol}}(t,\tau,\vec{p},\vec{p}\,') =  
A(\vec{p},\vec{p}\,') 
\Big[ G_e(Q^2) M_N \big( E_N(\vec{p}\,) + E_N(\vec{p}\,')\big) 
\\ {} \times
\big( \vec{p}\,' \cdot \vec{p} 
     - (M_N + E_N(\vec{p}\,)) (M_N + E_N(\vec{p}\,')) \big)
\\  {}
+ G_m(Q^2) \big( (\vec{p}\,' \cdot \vec{p}\,)^2 
          - \vec{p}\,^2 \vec{p}\,^{\prime 2} \big) \Big] \,, 
\end{multline} 
\begin{multline} \label{r4pol}
R_4^{\mathrm {pol}}(t,\tau,\vec{p},\vec{p}\,')  =   
A(\vec{p},\vec{p}\,') 
\mathrm i (p_1' p_3 - p_3' p_1) 
\Big[ G_e(Q^2) M_N \big( E_N(\vec{p}\,) + E_N(\vec{p}\,')\big)
\\ {}
 +   G_m(Q^2) \big( \vec{p}\,' \cdot \vec{p}
     - (M_N + E_N(\vec{p}\,)) (M_N + E_N(\vec{p}\,')) \big) \Big] \,,
\end{multline} 
and for $j=1,2,3$
\begin{multline} \label{rjunpol}
 R_j^{\mathrm {unpol}}(t,\tau,\vec{p},\vec{p}\,') =  
A(\vec{p},\vec{p}\,') \mathrm i
\Big[ G_e(Q^2) M_N \big( p_j + p_j'\big)
\big( (M_N + E_N(\vec{p}\,)) (M_N + E_N(\vec{p}\,')) 
           - \vec{p}\,' \cdot \vec{p}\big)
\\ {}
+ G_m(Q^2) \big( p_j ( E_N(\vec{p}\,) \vec{p}\,^{\prime 2}
 - E_N(\vec{p}\,') \vec{p}\,' \cdot \vec{p}\, )
+  p_j' ( E_N(\vec{p}\,') \vec{p}\,^2
 - E_N(\vec{p}\,) \vec{p}\,' \cdot \vec{p}\, ) \big) \Big] \,, 
\end{multline} 
\begin{multline} \label{rjpol} 
R_j^{\mathrm {pol}}(t,\tau,\vec{p},\vec{p}\,') =  
A(\vec{p},\vec{p}\,') 
\Big[ G_e(Q^2) M_N \big( p_j + p_j'\big)
\big( p_1' p_3 - p_3' p_1 \big)
+ G_m(Q^2) \big( M_N (p_2 + p_2') (\vec{p}\,' \times \vec{p})_j
\\
{} + \big( (M_N + E_N(\vec{p}\,)) (M_N + E_N(\vec{p}\,')) 
           - \vec{p}\,' \cdot \vec{p} \big)
\sum_{k=1}^3 \epsilon_{j2k} ( p_k' E_N(\vec{p}\,) - p_k E_N(\vec{p}\,'))
 \big) \Big] \,. 
\end{multline} 
Analogous expressions for the computation of the form factors 
$F_1$ and $F_2$ are obtained by inserting the definitions (\ref{sachs}) 
in the above equations.

\begin{figure}
  \includegraphics[width=14.0cm]{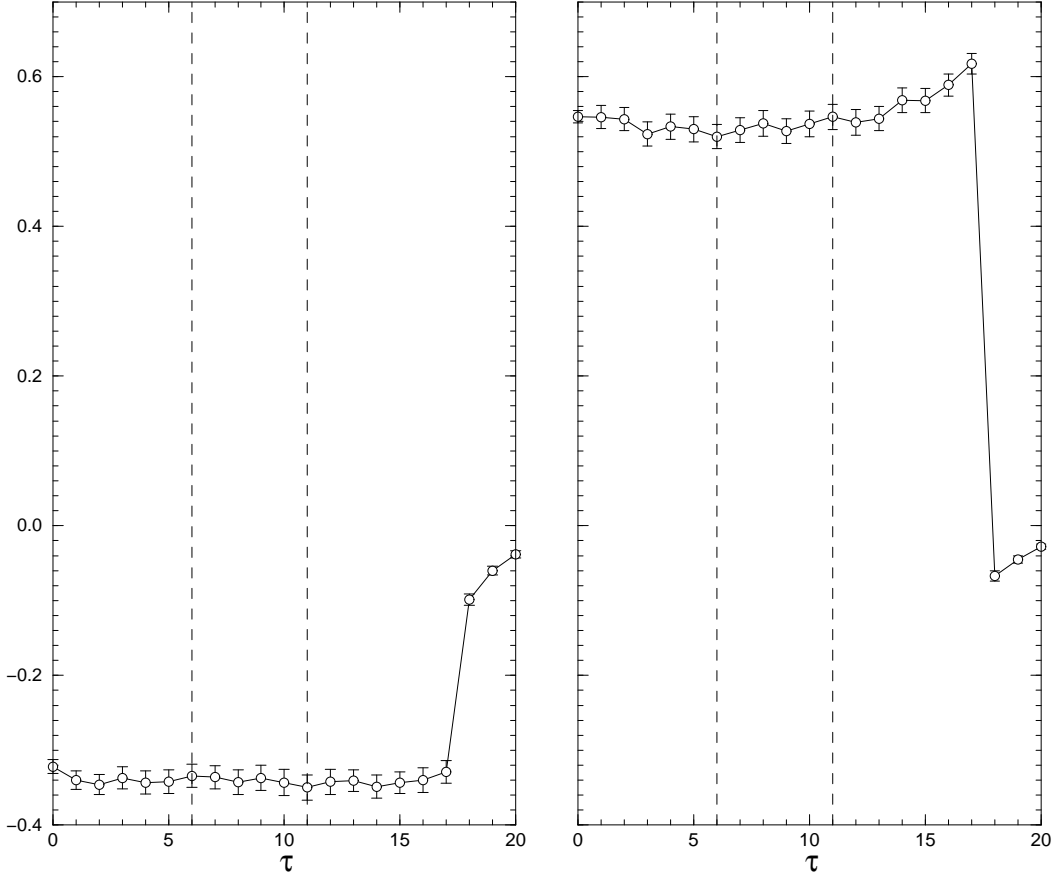}
    \caption{The ratios $R_3^{\mathrm {pol}}$ (left) and 
             $R_4^{\mathrm {unpol}}$ (right) plotted versus $\tau$
             for $\beta = 6.2$, $\kappa = 0.1344$. Here $\vec{p} = \vec{0}$ 
             and $\vec{q}$ is the fourth momentum in the list (\ref{qlist}).  
             The vertical dashed lines indicate the fit range for the
             extraction of the plateau value.}
    \label{fig:plateau}
\end{figure}

We have computed the ratios $R$ in (\ref{ratio}) for two choices of 
the momentum $\vec{p}$,
\begin{equation} 
\frac{L}{2 \pi} \vec{p} = 
\left ( \begin{array}{r} 0 \\ 0 \\ 0 \end{array} \right) \,,\,
\left ( \begin{array}{r} 1 \\ 0 \\ 0 \end{array} \right) \,,
\end{equation}
and eight choices of the vector $\vec{q} = \vec{p}\,' - \vec{p}$,
\begin{equation} \label{qlist}
\frac{L}{2 \pi} \vec{q}  =  
\left ( \begin{array}{r} 0 \\ 0 \\ 0 \end{array} \right) \,,\,
\left ( \begin{array}{r} 0 \\ -1 \\ 0 \end{array} \right) \,,\,
\left ( \begin{array}{r} 0 \\ -2 \\ 0 \end{array} \right) \,,\,
\left ( \begin{array}{r} -1 \\ 0 \\ 0 \end{array} \right) \,,\,
\left ( \begin{array}{r} -2 \\ 0 \\ 0 \end{array} \right) \,,\,
\left ( \begin{array}{r} -1 \\ -1 \\ 0 \end{array} \right) \,,\,
\left ( \begin{array}{r} -1 \\ -1 \\ -1 \end{array} \right) \,,\,
\left ( \begin{array}{r} 0 \\ 0 \\ -1 \end{array} \right) \,,
\end{equation}
where $L$ denotes the spatial extent of the lattice.
In Fig.~\ref{fig:plateau} we show two examples of these ratios plotted
versus $\tau$ (for the unimproved operator). The final results for $R$ have
been determined by a fit with a constant in a suitable $\tau$ interval.
The corresponding errors have been computed by a jackknife procedure.
The values chosen for $t$ and for the fit intervals are collected in 
Table~\ref{tab:fits}.
\begin{table}
\caption{Sink positions $t$ and fit intervals (in $\tau$) used for the 
         extraction of the ratios $R$.}
\label{tab:fits}
\begin{ruledtabular}
\begin{tabular}{cccc}
$\beta$   & $t$  & fit interval \\ 
\hline 
6.0       & 13     & [4,9] \\
6.2       & 17     & [6,11] \\
6.4       & 23     & [7,16]
\end{tabular}
\end{ruledtabular}
\end{table}

Generically, several combinations of the above momenta lead to the same
$Q^2$, and several ratios $R$ contain the form factors at this $Q^2$
with non-vanishing coefficients. Hence we determined $G_e(Q^2)$ and 
$G_m(Q^2)$ from an (uncorrelated) MINUIT fit of all these 
$R$s with the corresponding expressions (\ref{r4unpol}) - (\ref{rjpol})
omitting all data points where the error for $R$ was larger than 25\%.
The results are collected in the tables in Appendix~B. 
A missing entry indicates a case where the corresponding form factor
could not be extracted, e.g.\ because we did not have sufficiently 
many $R$s with less than 25\% error.

The nucleon masses used can be found in Table~\ref{tab:param}. The 
corresponding errors were, however, not taken into account when computing
the errors of the form factors. Varying the nucleon masses within one
standard deviation changed the form factors only by fractions of the quoted 
statistical error.

In general, the nucleon three-point functions consist of a quark-line 
connected contribution and a quark-line disconnected piece. Unfortunately, 
the quark-line disconnected piece is very hard to compute (for some recent
attempts see Refs.~\cite{disco1,disco2,wilcox}). 
Therefore it is usually neglected, 
leading to one more source of systematic uncertainty.
However, in the case of exact isospin invariance the disconnected 
contribution drops out in non-singlet quantities like the isovector
form factors. That is why the isovector form factors 
(Tables~\ref{tab:clover60isov} - \ref{tab:clover64isov}) are our favorite
observables. Nevertheless, we have also computed the proton
form factors separately ignoring the disconnected contributions. The
results are given in
Tables~\ref{tab:clover60prot} - \ref{tab:clover64prot}.
Regrettably, meaningful values of the electric form factor of the
neutron could not be extracted from our data
(for a more successful attempt see Ref.~\cite{tang}).
The results for the neutron magnetic form factor are collected in 
Tables~\ref{tab:clover60neut} - \ref{tab:clover64neut}. 
Note that the isovector form factors have been computed directly 
(cf.\ Eq.~(\ref{isovector})) and not as the difference of the proton 
and neutron form factors. 

\section{Chiral extrapolation: a first attempt}
\label{sect:first}

The quark masses in our simulations are considerably
larger than in reality leading to pion masses above 500 MeV.
Hence we cannot compare our results with 
experimental data without performing a chiral extrapolation. 
In a first analysis of the proton results 
(see Refs.~\cite{lat98,adelaide}) we assumed a linear quark-mass dependence
of the form factors. More precisely, we proceeded as follows.

Schematically, the relation between a ratio $R$ 
(three-point function/two-point function) and the form factors $G_e$, $G_m$
can be written in the form
\begin{equation} 
R = \langle p' | J | p \rangle + \cdots 
 = c_e G_e + c_m G_m
\end{equation}
with known coefficients $c_e$, $c_m$ for each data point characterized
by the momenta, the quark mass, the spin projection and 
the space-time component of the current. 
Assuming a linear quark-mass dependence of $G_e$ and $G_m$ 
we performed a 4-parameter fit,
\begin{equation} 
R = c_e a_e (a m_q) + c_e b_e +  c_m a_m (a m_q) + c_m b_m \,,
\end{equation}
of all ratios $R$ belonging to the same value of $Q^2$ in the chiral limit.
The resulting form factors in the chiral limit are typically larger than
the experimental data. They can be fitted with a dipole
form, but the masses from these fits are considerably larger than their
phenomenological counterparts~\cite{lat98,adelaide}.

What could be the reason for this discrepancy? Several possibilities 
suggest themselves: finite-size effects, quenching errors, cut-off effects
or uncertainties in the chiral extrapolation. The length $L$ of the 
spatial boxes in our simulations is such that the inequality 
$m_\pi L > 4 $ holds in all cases. 
Previous experience suggests that in the quenched approximation 
this is sufficient to exclude 
considerable distortions of the results due to the finite volume.
This assumption is confirmed by simulations with Wilson 
fermions, where we have data on different volumes. 
Quenching errors are much more difficult to control. However, first
simulations with dynamical fermions indicate that -- for the rather 
heavy quarks we can deal with -- the form factors do not change very much
upon unquenching~\cite{adelaide}. Having Monte Carlo data for 
three different lattice spacings (see Table~\ref{tab:param}) we 
can test for cut-off effects in the chirally extrapolated form 
factors, but we find them to be hardly significant. So our 
chiral extrapolation ought to be reconsidered. Indeed, the chiral 
extrapolation of lattice data has been discussed intensively in 
the recent literature (see, e.g., 
Refs.~\cite{wilcox,tang,chiextra,chiralmag,australia1a,australia1b,
australia2,australia3})
and it has been pointed out that the issue is highly non-trivial.
Therefore we shall examine the quark-mass dependence of our
form factors in more detail.

Ideally, one would like to identify a regime of parameters (quark masses
in particular) where contact with chiral effective field theory (ChEFT) 
can be made on the basis of results like those presented for the nucleon form
factors in Ref.~\cite{chpt}.
Once the range of applicability of these low-energy effective 
field theories has been established,
one can use them for a safe extrapolation to smaller masses. 
However, these schemes do not work for arbitrarily
large quark masses (or pion masses), nor for arbitrarily large 
values of $Q^2$. In particular, the expressions for the form factors 
worked out in Ref.~\cite{chpt} can be trusted only up to 
$Q^2 \approx 0.4 \, \text{GeV}^2$ (see the discussion in 
Sec.~\ref{sect:chpt.ff} below) . Unfortunately, from our 
lattice simulations we only have data for 
values of $Q^2$ which barely touch the interval 
$0 < Q^2 < 0.4 \, \text{GeV}^2$. Therefore we shall try to describe the $Q^2$
dependence of the lattice data for each quark mass by a suitable ansatz
(of dipole type) 
and then study the mass dependence of the corresponding parameters. The
fit ansatz will also serve as an extrapolation of the magnetic form 
factor down to $Q^2=0$. Since we cannot compute
$G_m(0)$ directly, such an extrapolation is required anyway to 
determine the magnetic moment. (For a different method, which does
not require an extrapolation, see Ref.~\cite{ji}.) In 
Sec.~\ref{sect:disc} we shall come back to a comparison with ChEFT.

\section{Investigating the quark-mass dependence}
\label{sect:invest}

The analysis of our form factor data sketched in Sec.~\ref{sect:first} 
yielded results in the chiral limit without much control
over the approach to that limit. 
In this section we want to study the quark-mass dependence of the 
form factors more thoroughly. As already mentioned, to this end
we have to make use of a suitable description of the $Q^2$ dependence.

Motivated by the fact that the experimentally measured form factors 
at small values of $Q^2$ can be described by a dipole form 
(cf.\ Eq.~(\ref{dipole})) we fitted our data with the ansatz
\begin{eqnarray} 
G_l (Q^2) & = & \frac{A_l}{(1 + Q^2/M_l^2)^2} \quad , \quad l=e,m \,, 
\nonumber \\ 
F_i (Q^2) & = & \frac{A_i}{(1 + Q^2/M_i^2)^2} \quad , \quad i=1,2 \,.
\end{eqnarray}
In the case of the form factors $G_e$ ($F_1$) we fixed $A_e=1$ ($A_1=1$).
Note that we do not require the dipole masses in the two form factors to
coincide. Thus our ansatz can accomodate deviations of the ratio
$G_m (0) G_e(Q^2) / G_m (Q^2)$ from unity as they have been observed 
in recent experiments~\cite{gao,JLAB1,JLAB2}. 

\begin{figure}
  \includegraphics[width=12.0cm]{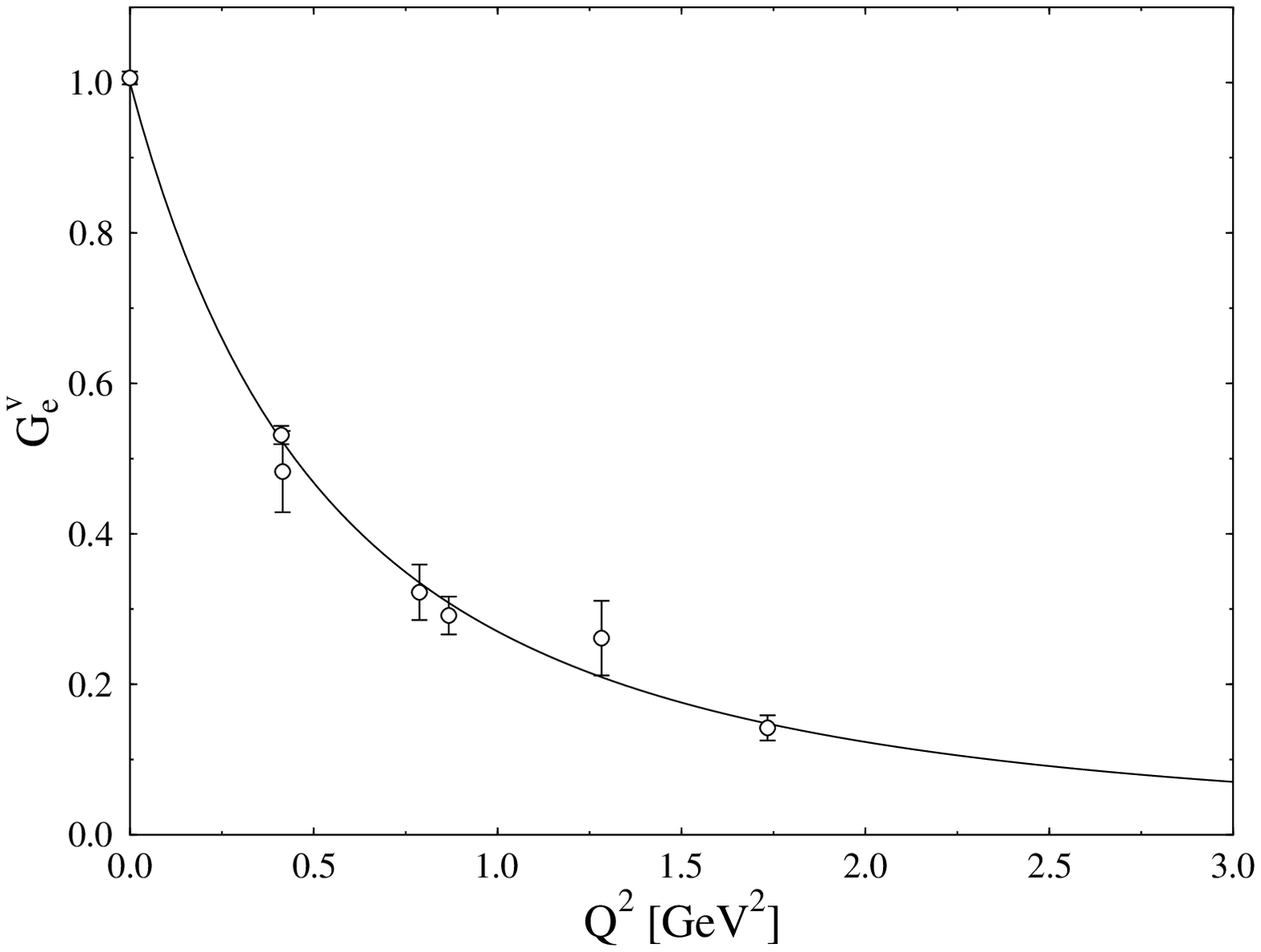} \\
  \includegraphics[width=12.0cm]{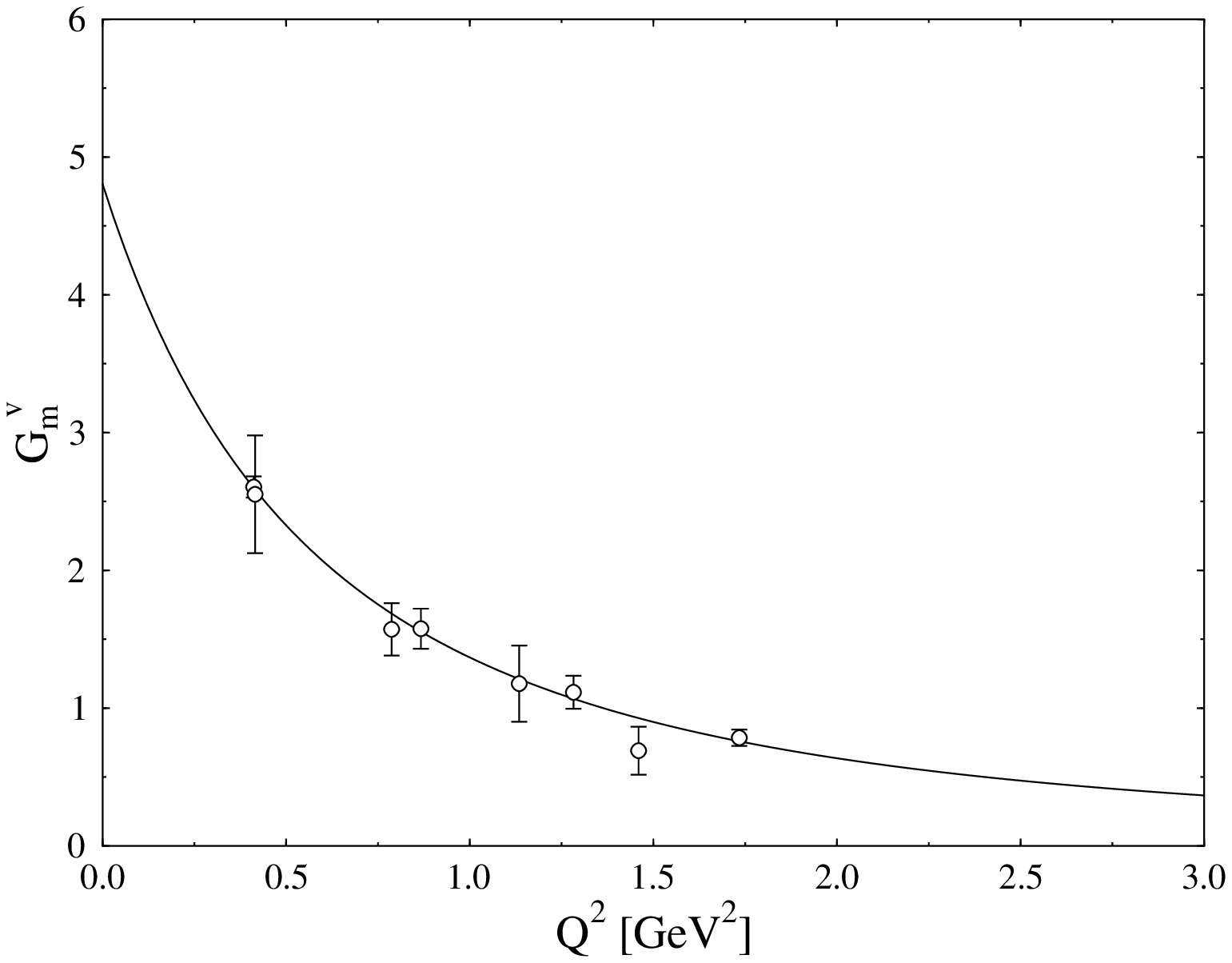} 
    \caption{Dipole fits of $G_e^v$ data (top) and $G_m^v$ data (bottom)
             at $\beta=6.4$ and $m_\pi = 0.648 \, \text{GeV}$.}
    \label{fig:dipfit}
\end{figure}

Indeed, for all masses considered in our simulations
the lattice data can be described rather well by a dipole ansatz. 
In Fig.~\ref{fig:dipfit} we show examples of our data (for 
$m_\pi = 0.648 \, \text{GeV}$) together with the 
dipole fits. The fit results are collected in Table~% 
\ref{tab:clover.dipisov} for the isovector
form factors, in Table~\ref{tab:clover.dipprot} for the proton 
form factors and in Table~\ref{tab:clover.dipneut} for the magnetic 
form factor of the neutron.

\begin{figure}
  \includegraphics[width=12.0cm]{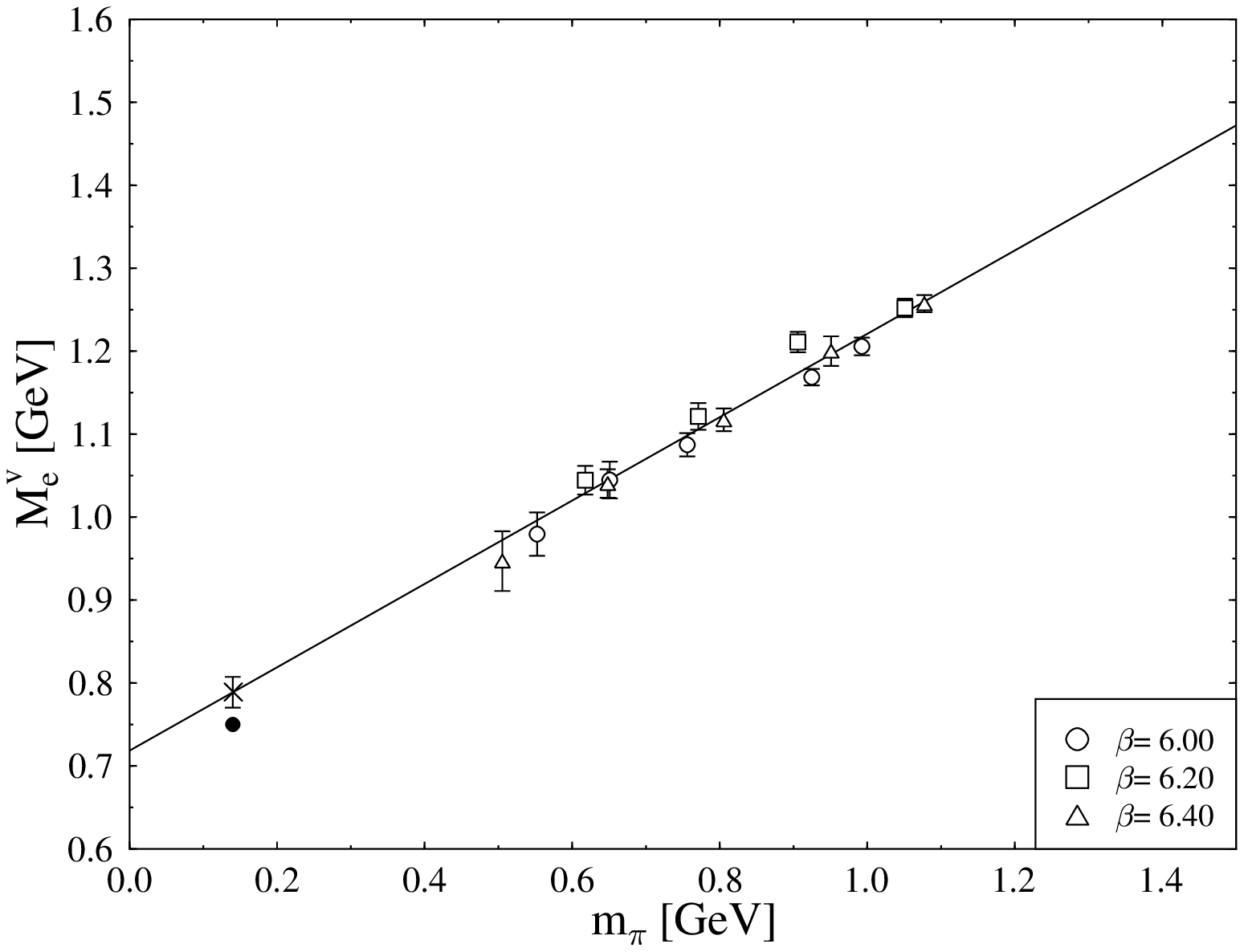} \\
  \includegraphics[width=12.0cm]{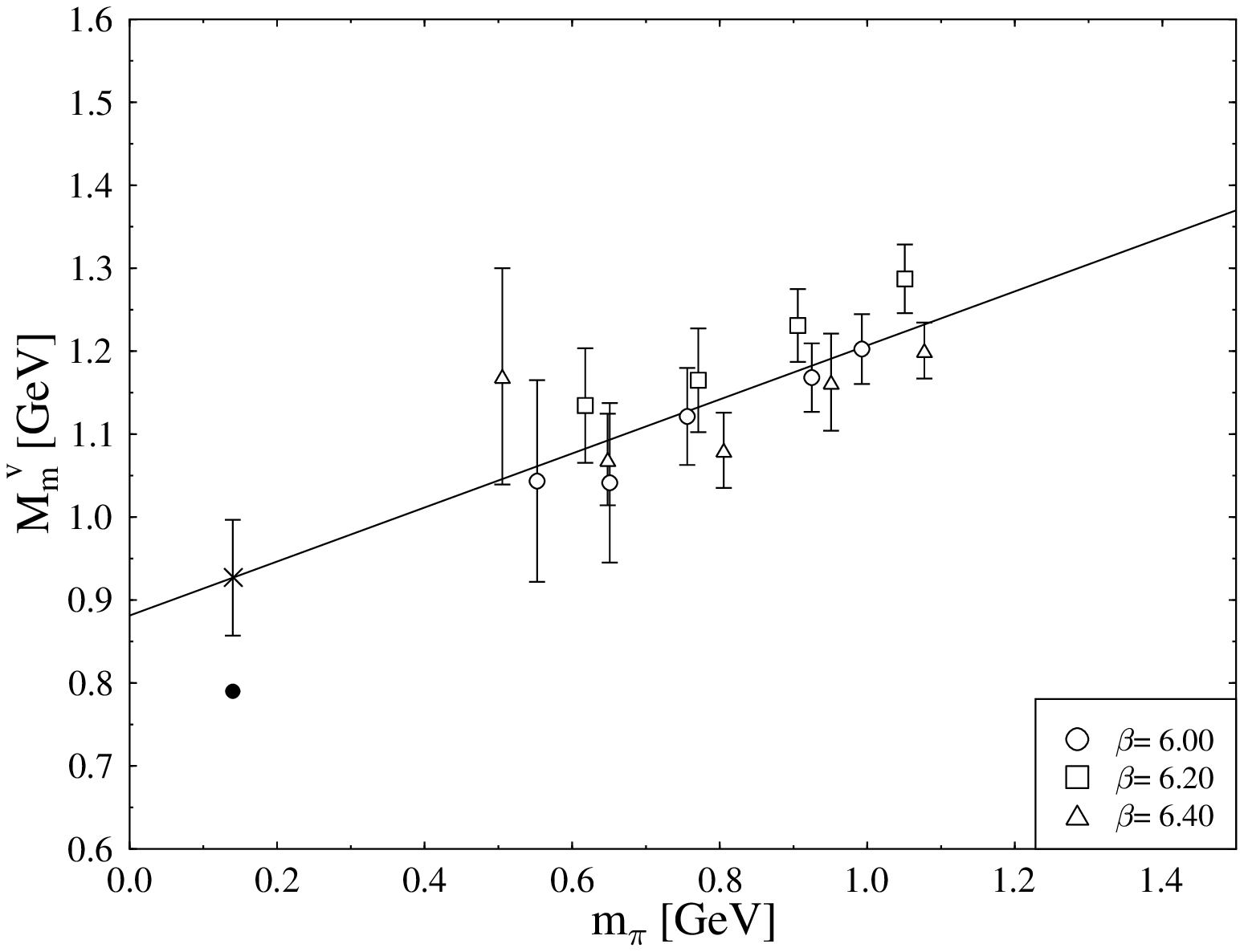}
    \caption{Isovector dipole masses together with linear fits.
             The extrapolated value at the physical pion mass is
              marked by a cross. The solid circle indicates the
             phenomenological value computed from the radii given 
             in Ref.~\cite{mergell}.}
    \label{fig:lfitmass}
\end{figure}

\begin{figure}
  \includegraphics[width=12.0cm]{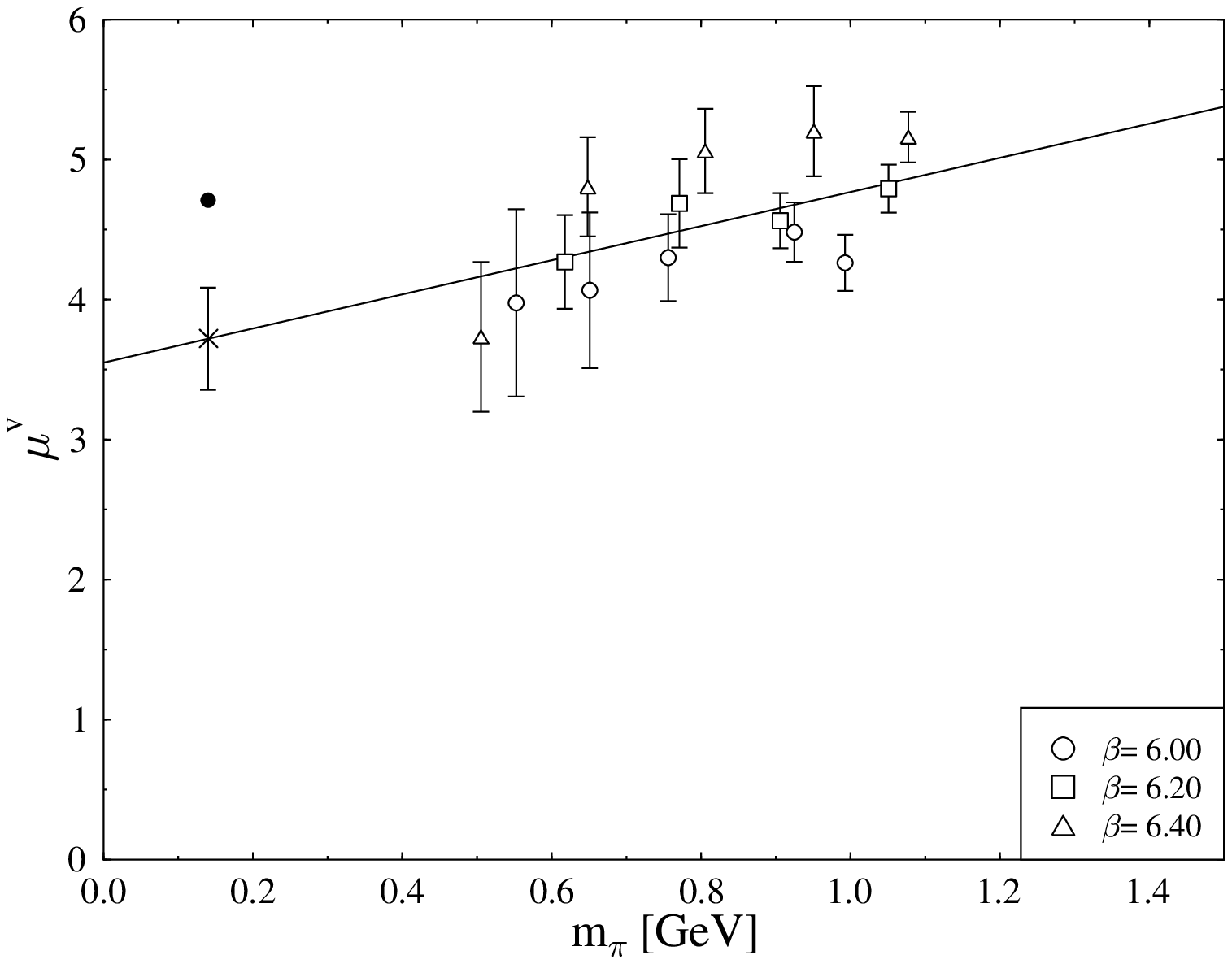}
    \caption{Isovector magnetic moment together with a linear fit.
             The extrapolated value at the physical pion mass is
              marked by a cross. The solid circle indicates the
             experimental value.}
    \label{fig:lfitmom}
\end{figure}

In Figs.~\ref{fig:lfitmass}, \ref{fig:lfitmom}
we plot the isovector electric dipole mass $M_e^v$, the isovector
magnetic dipole mass $M_m^v$ and the isovector magnetic moment $\mu^v$
(extracted from the Sachs form factors)
versus $m_\pi$. We make the following observations. 

Scaling violations in the dipole masses seem to be smaller than the
statistical accuracy, since the results do not show a definite 
trend as $\beta$ grows from 6.0 to 6.4.  
For the magnetic moments the situation is less clear. There might be
some systematic shift, though not much larger than the statistical errors.

The electric dipole masses tend to become slightly smaller than the 
magnetic dipole masses as the pion mass decreases though it is not 
clear whether this difference is statistically significant. 
This behavior agrees qualitatively with the recent JLAB 
data~\cite{JLAB1,JLAB2} for $G_e/G_m$ in the proton 
(see Fig.~\ref{fig:ratio} below).

The data for the electric dipole masses suggest a linear 
dependence on $m_\pi$. Therefore we could not resist temptation 
to perform linear fits of the dipole masses and moments 
(extracted from the Sachs form factors)
in Tables~\ref{tab:clover.dipisov} - \ref{tab:clover.dipneut} in order to
obtain values at the physical pion mass. Of course, at some point the
singularities and non-analyticities arising from the Goldstone
bosons of QCD must show up and will in some observables lead to a 
departure from the simple linear behavior. It is however conceivable 
that this happens only at rather small pion masses (perhaps even only 
below the physical pion mass) and thus does not influence the value at the 
physical pion mass too strongly. 
One can try to combine the leading non-analytic behavior of chiral
perturbation theory with a linear dependence on $m_\pi^2$ as it is
expected at large quark mass in order to obtain an interpolation formula
valid both at small and at large masses. Fitting our form factor data 
with such a formula one ends up remarkably close to the experimental
results~\cite{australia3}.

We performed our fits separately for each $\beta$ value as well as for 
the combined data from all three $\beta$ values. 
The results are 
presented in Table~\ref{tab:linfits} together with the experimental 
numbers. For the isovector dipole masses and the isovector magnetic 
moment the fit curves (from the joint fits for all $\beta$ values) 
are plotted in Figs.~\ref{fig:lfitmass}, \ref{fig:lfitmom}. The 
corresponding plots for the proton and neutron data look similar.
In the case of the electric dipole mass, the extrapolated result
lies remarkably close to the experimental value. For the magnetic
dipole mass and the magnetic moment
the agreement is less good, but still satisfactory
in view of the relatively large statistical errors.

\begin{table*}
\caption{Results at the physical pion mass from linear fits, 
         separately for each $\beta$ value as well as for the combined data. 
         The experimental numbers for $M_e^v$ and $M_m^v$ were derived 
         from the radii given in Ref.~\cite{mergell} 
         (cf.\ Eq.~(\ref{dipm}) below).}
\label{tab:linfits}
\begin{ruledtabular}
\begin{tabular}{cddddd}
 {} & \multicolumn{1}{c}{$\beta = 6.0$} & 
 \multicolumn{1}{c}{$\beta = 6.2$} & \multicolumn{1}{c}{$\beta = 6.4$} & 
\multicolumn{1}{c}{combined} & \multicolumn{1}{c}{experiment}   \\
\hline
$ M_e^v $ [GeV]      & 0.78(3)  & 0.82(3)  & 0.77(3)  & 0.789(10) & 0.75 \\
$ M_m^v $ [GeV]      & 0.87(15) & 0.94(13) & 0.93(10) & 0.93(7)   & 0.79 \\
$ \mu^v $            & 3.9(8)   & 3.9(6)   & 3.9(6)   & 3.7(4)    & 4.71 \\
$ M_e^p $ [GeV]      & 0.80(3)  & 0.84(3)  & 0.80(2)  & 0.807(15) & 0.84 \\ 
$ M_m^p $ [GeV]      & 0.93(15) & 0.94(13) & 0.92(10) & 0.93(7)   & 0.84 \\
$ \mu^p $            & 2.3(5)   & 2.4(4)   & 2.4(3)   & 2.3(2)    & 2.79 \\
$ M_m^n $ [GeV]      & 0.83(16) & 0.88(15) & 0.91(11) & 0.89(8)   & 0.84 \\
$ \mu^n $            & -1.6(4)  & -1.6(3)  & -1.5(3)  & -1.47(17) & -1.91
\end{tabular}
\end{ruledtabular}
\end{table*}

Using the dipole approximations of the proton form factors 
with the extrapolated dipole masses as given in the 
fifth column of Table~\ref{tab:linfits} we can now compare 
\begin{equation} 
\mu^p \frac{G_e^p(Q^2)}{G_m^p (Q^2)} = 
 \frac{\left(1 + Q^2/(M_m^p)^2 \right)^2}
      {\left(1 + Q^2/(M_e^p)^2 \right)^2}
\end{equation}
with the experimental data from Refs.~\cite{JLAB1,JLAB2}. This is done in 
Fig.~\ref{fig:ratio}. 
Especially for the larger values of $Q^2$ we find good agreement,
although the lattice data only cover the range 
$Q^2 < 2 \, \text{GeV}^2$ and for $Q^2 > 2 \, \text{GeV}^2$ the curve
represents an extrapolation. 
It is perhaps not too surprising that the agreement improves as $Q^2$ grows:
Larger $Q^2$ probe smaller distances inside the proton where
the influence of the pion cloud, which is only insufficiently
taken into account in the quenched approximation, is diminished.

\begin{figure}
  \includegraphics[width=12.0cm]{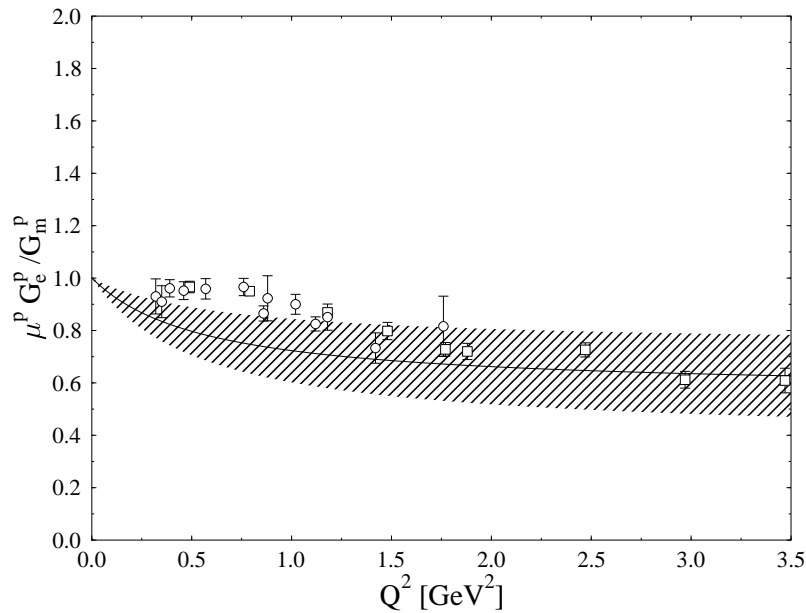}
    \caption{The ratio $\mu^p G_e^p / G_m^p$ from the chirally
             extrapolated dipole fits of the proton form factors
             (curve) compared with the experimental data from 
             Refs.~\cite{JLAB1} (squares) and \cite{JLAB2} (circles).
             The error band (indicated by the hatched area) has been 
             computed from the errors of the
             extrapolated dipole masses. For the experimental numbers 
             only the statistical errors are shown.} 
    \label{fig:ratio}
\end{figure}

\section{Results from chiral effective field theory}
\label{sect:chpt}

\subsection{From diagrams to form factors}
\label{sect:chpt.ff}

\begin{figure}
  \includegraphics[width=12.0cm]{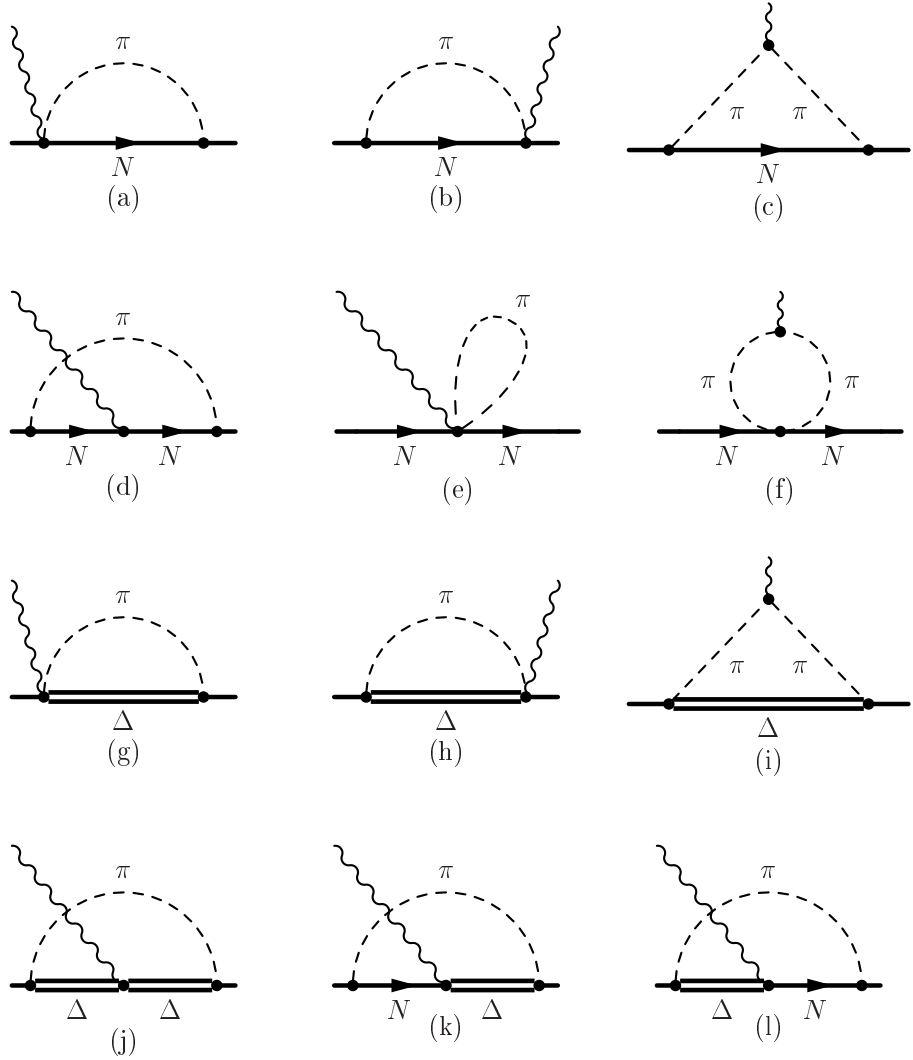}
    \caption{One-loop diagrams in SSE contributing to the electromagnetic 
             form factors. The wiggly line represents an external vector
             field.}
    \label{fig:diag}
\end{figure}

For the comparison with ChEFT we choose the isovector
form factors, because they do not suffer from the problem of quark-line 
disconnected contributions.
Recently, a calculation for the quark-mass dependence of the isovector 
anomalous magnetic moment has been presented~\cite{chiralmag}. The authors 
employed a ChEFT with explicit nucleon and $\Delta$ degrees of freedom, 
called the Small Scale Expansion (SSE)~\cite{review}. It was argued 
\cite{chiralmag} that the standard power-counting of ChEFT had to be changed 
to obtain a well-behaved chiral expansion -- in particular, the 
leading isovector $N\Delta$ transition coupling $c_V$ (not to be confused
with the improvement coefficient used earlier) had to be 
included in the leading-order Lagrangian. For details we refer 
to Ref.~\cite{chiralmag}. 
The formula for the nucleon mass obtained in this framework is
given in Eq.~(\ref{nuclmass}).
Here we extend this analysis from
the magnetic moments to the Dirac and Pauli form factors of the nucleon, 
utilizing the same Lagrangians and couplings as in~\cite{chiralmag}. 
To leading one-loop order (${\mathcal O}(\epsilon^3)$ in SSE) 12 diagrams
shown in Fig.~\ref{fig:diag} have to be evaluated in addition to the 
short-distance contributions. 
The calculation follows very closely the one presented in 
Ref.~\cite{chpt}, where further technical details of form factor 
calculations in ChEFT are discussed. The main difference between 
our analysis here and Ref.~\cite{chpt} arises from the modified
counting of $c_V$, leading to the additional diagrams (k) and (l)
in Fig.~\ref{fig:diag}. 
Evaluating all diagrams in the Breit frame, we identify the isovector 
form factors $F_1^v(q^2)$ and $F_2^v(q^2)$ via the ${\mathcal O}(\epsilon^3)$ 
relation for the proton matrix element
\begin{multline} \label{match}
\langle p_2|\left( \bar{u} \gamma_\mu u - \bar{d} \gamma_\mu d \right)
|p_1\rangle_{\mathrm {Breit}}
= \frac{e}{N_1 N_2}\;\bar{u}_v(r_2)\left[\left(F_1^{\;v}(q^2)+
         \frac{q^2}{4 (M_N^0)^2}
   F_2^{\;v}(0)+{\mathcal O}(\epsilon^4)\right)v_\mu\right.
\\
 \left.{} +\frac{1}{M_N^0}\left(F_1^{\;v}(0)+F_2^{\;v}(q^2)+
     {\mathcal O}(\epsilon^4)\right)
   \left[S_\mu,S_\nu\right] q^\nu\right]u_v(r_1) 
\end{multline}
written in Minkowski space notation.
Here $M_N^0$ is the nucleon mass in the chiral limit 
and $u_v(r_i)$ denotes a nucleon spinor with the normalization 
$N_i$ as given in Ref.~\cite{chpt}.
The quantity $S_\mu$ denotes the Pauli-Lubanski spin-vector, 
$S_\mu = \frac{\mathrm i}{2} \gamma_5 \sigma_{\mu \nu} v^\nu$. 
The four-vector $v^\mu$ ($v^2=1$) is connected to the usual four-momentum 
vector $p^\mu=M_N^0 v^\mu+r^\mu$, where 
$r_\mu$ is a soft momentum. 
Further details regarding calculations 
in this non-relativistic ChEFT can be found in Ref.~\cite{review}.

Nevertheless we have to discuss some subtleties behind Eq.~(\ref{match}) 
to be able to compare the ChEFT results to lattice data. 
In (lattice) QCD a change in the quark mass
does not only lead to a change in  $\mu^v$ and $\kappa_v$, but at the same
time also to a change in the nucleon mass. However, this variation of
the nucleon mass -- corresponding to a quark-mass dependent ``magneton'' -- 
is not accounted for at the order in ChEFT we are working at, as can
be seen from the presence of the nucleon mass in the chiral limit 
$M_N^0$ in Eq.~(\ref{match}). Hence, before comparing with the lattice
results, we have to eliminate this effect.
We do so by defining ``normalized'' magnetic moments which are measured
relative to the {\em physical} mass of the nucleon $M_N^{\mathrm {phys}}$
and so are given in units of ``physical'' magnetons. These normalized 
magnetic moments can then be matched to the formulae from ChEFT with
$M_N^0$ replaced by $M_N^{\mathrm {phys}}$.

We define the normalized magnetic moment by
\begin{equation} \label{muvmatch}
\mu^v_{\mathrm {norm} } := \mu^v_{\mathrm {lattice}}\; 
   \frac{M_N^{\mathrm {phys}}}{M_N^{\mathrm {lattice}}} 
   = \frac{M_N^{\mathrm {phys}}}{M_N^{\mathrm {lattice}}} 
     + \kappa_v^{\mathrm {lattice}}\;\frac{M_N^{\mathrm {phys}}}
                       {M_N^{\mathrm {lattice}}} \,,
\end{equation}
Correspondingly, we take for the normalized anomalous magnetic moment 
\begin{equation}
\kappa_v^{\mathrm {norm}} :=  \kappa_v^{\mathrm {lattice}}\; 
\frac{M_N^{\mathrm {phys}}}{M_N^{\mathrm {lattice}}}
\end{equation}
such that 
\begin{equation}
\mu^v_{\mathrm {norm}} = 
\frac{M_N^{\mathrm {phys}}}{M_N^{\mathrm {lattice}}} + 
\kappa_v^{\mathrm {norm}} \,.
\end{equation}
At higher orders in the chiral 
expansion, the quark-mass dependence of the nucleon mass will 
manifest itself in the matrix element (\ref{match}), and
the fits will have to be modified accordingly.

\subsection{Form factors at $\mathcal O (\epsilon^3)$}

For the isovector Dirac form factor we obtain
\begin{eqnarray} 
{} & {} & 
F_{1}^{v}(q^2) =  1 + \frac{1}{(4\pi F_\pi)^2}
                   \left\{ q^2 \left(\frac{68}{81} c_A^2 
           - \frac{2}{3}g_{A}^2-2 B_{10}^{(r)}(\lambda) \right) \right. 
   + q^2 \left(\frac{40}{27} c_A^2-\frac{5}{3}g_{A}^2
   -\frac{1}{3}\right)
                   \log\left[\frac{m_\pi}{\lambda}\right]  
\nonumber \\  {} & {} & 
     {} + \int_{0}^{1}dx \left[\frac{16}{3}\Delta^2 c_A^2
          + m_{\pi}^2 \left(3 g_{A}^2+1-\frac{8}{3} c_A^2 \right) \right.
        - \left. q^2 x(1-x)\left(5 g_{A}^2+1
        -\frac{40}{9}c_A^2\right)\right] 
        \log\left[\frac{\tilde{m}^2}{m_{\pi}^2}\right]  
\nonumber \\ {} & {} & 
     {} + \int_{0}^{1}dx \left[ \frac{32}{9} c_A^2 q^2 x(1-x) 
       \frac{\Delta \log R(\tilde{m})}{\sqrt{\Delta^2-\tilde{m}^2}} \right] 
\nonumber \\  {} & {} & {}
    -  \left. \int_{0}^{1}dx \; \frac{32}{3}c_A^2 \Delta 
      \left[\sqrt{\Delta^2-m_{\pi}^2}\log R(m_\pi) 
      -\sqrt{\Delta^2-\tilde{m}^2}\log R(\tilde{m}) \right] \right\} 
      + \mathcal O (\epsilon^4) \,. 
\label{F1}
\end{eqnarray}
To the same order one finds
\begin{multline}  \label{F2}
 F_{2}^{v}(q^2) =  \kappa_v (m_\pi) - g_{A}^2
           \frac{4\pi M_N}{(4\pi F_\pi)^2} 
           \int_{0}^{1}dx\left[ \sqrt{\tilde{m}^2}-m_{\pi}\right]
\\
         {} +\frac{32c_A^2 M_N \Delta}{9 (4\pi F_\pi)^2}\int_{0}^{1}dx 
         \left[ \frac{1}{2}\log\left[\frac{\tilde{m}^2}{4\Delta^2}\right]
         -\log\left[\frac{m_\pi}{2\Delta}\right] \right.  
\\
  \left. {} +\frac{\sqrt{\Delta^2-\tilde{m}^2}}{\Delta} \log R(\tilde{m})
              -\frac{\sqrt{\Delta^2-m_{\pi}^2}}{\Delta} \log R(m_\pi) \right] 
\end{multline}
for the isovector Pauli form factor,
where we have used the abbreviations
\begin{equation}
R(m)
= \frac{\Delta}{m}+\sqrt{\frac{\Delta^2}{m^2}-1} \,, \quad
\tilde{m}^2 = m_\pi^2 - q^2 x (1-x) \,.
\end{equation}
Furthermore, the isovector anomalous magnetic moment $\kappa_v (m_\pi)$
appearing in Eq.~(\ref{F2}) is given by      
\begin{multline} \label{kappa}
 \kappa_v (m_\pi) = \kappa_v^0-\frac{g_A^2\,m_\pi M_N}{4\pi F_\pi^2}
                +   \frac{2 c_A^2 \Delta M_N}{9\pi^2 F_\pi^2}
             \left\{\sqrt{1-\frac{m_\pi^2}{\Delta^2}}\log R(m_\pi)
           +\log\left[\frac{m_\pi}{2\Delta}\right] \right\}
\\
        -   8 E_1^{(r)} (\lambda) M_N m_\pi^2
        +  \frac{4c_A c_V g_A M_N m_\pi^2}{9\pi^2 F_\pi^2}
         \log\left[\frac{2\Delta}{\lambda} \right] 
        +  \frac{4c_A c_V g_A M_N m_\pi^3}{27\pi F_\pi^2\Delta}
\\
         -   \frac{8 c_A c_V g_A \Delta^2 M_N}{27\pi^2 F_\pi^2}
         \Bigg\{\left(1-\frac{m_\pi^2}{\Delta^2}\right)^{3/2} \log R(m_\pi)  
          + \left(1-\frac{3m_\pi^2}{2\Delta^2}\right)
           \log\left[\frac{m_\pi}{2\Delta}\right] \Bigg\}   
\end{multline}
to $\mathcal O (\epsilon^3)$. As already mentioned, 
to this order the nucleon mass $M_N$ is a fixed number in the above 
expression, independent of the quark mass, and we shall later 
identify it with $M_N^{\mathrm {phys}}$.
Note that Eq.~(\ref{kappa}) corresponds to case $C$ 
in the terminology of Ref.~\cite{chiralmag}. Of course, it
agrees with the result obtained in Ref.~\cite{chiralmag} because the 
magnetic moments are automatically contained
in a calculation of the form factors, as can be seen from the diagrams
of Fig.~\ref{fig:diag}.

The expressions (\ref{F1}) and (\ref{F2}) contain a number of 
phenomenological parameters:
the pion decay constant $F_\pi$, the leading axial 
$N \Delta$ coupling $c_A$ 
(denoted by $g_{\pi N \Delta}$ in Ref.~\cite{chpt}), the
axial coupling of the nucleon $g_A$, the nucleon mass $M_N$ and 
the $\Delta$(1232)-nucleon mass splitting $\Delta = M_\Delta - M_N$.
In addition, there is one parameter not directly related to phenomenology, 
$B_{10}^{(r)}(\lambda)$. This counterterm at the renormalization scale 
$\lambda$ parametrizes short-distance contributions to the Dirac radius
discussed in the next subsection. 
Further parameters are encountered in the expression 
(\ref{kappa}) for $\kappa_v (m_\pi)$: 
the isovector anomalous magnetic moment of the nucleon in the chiral limit 
$\kappa_v^0$, the leading isovector $N \Delta$ coupling $c_V$ and the
counterterm $E_1^{(r)} (\lambda)$, which leads to quark-mass dependent
short-distance contributions to $\kappa_v$.

\begin{figure}
  \includegraphics[width=12.0cm]{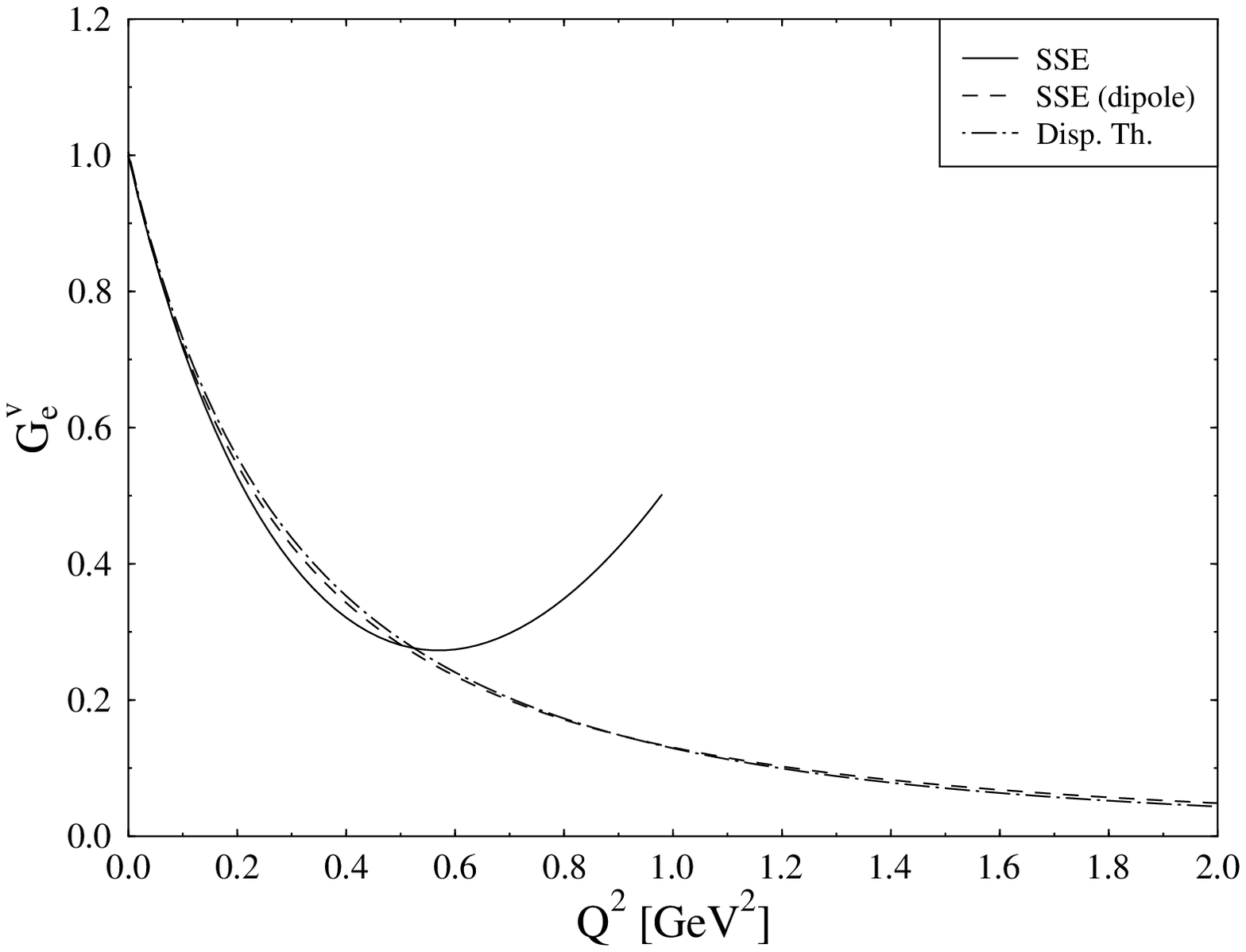} \\
  \includegraphics[width=12.0cm]{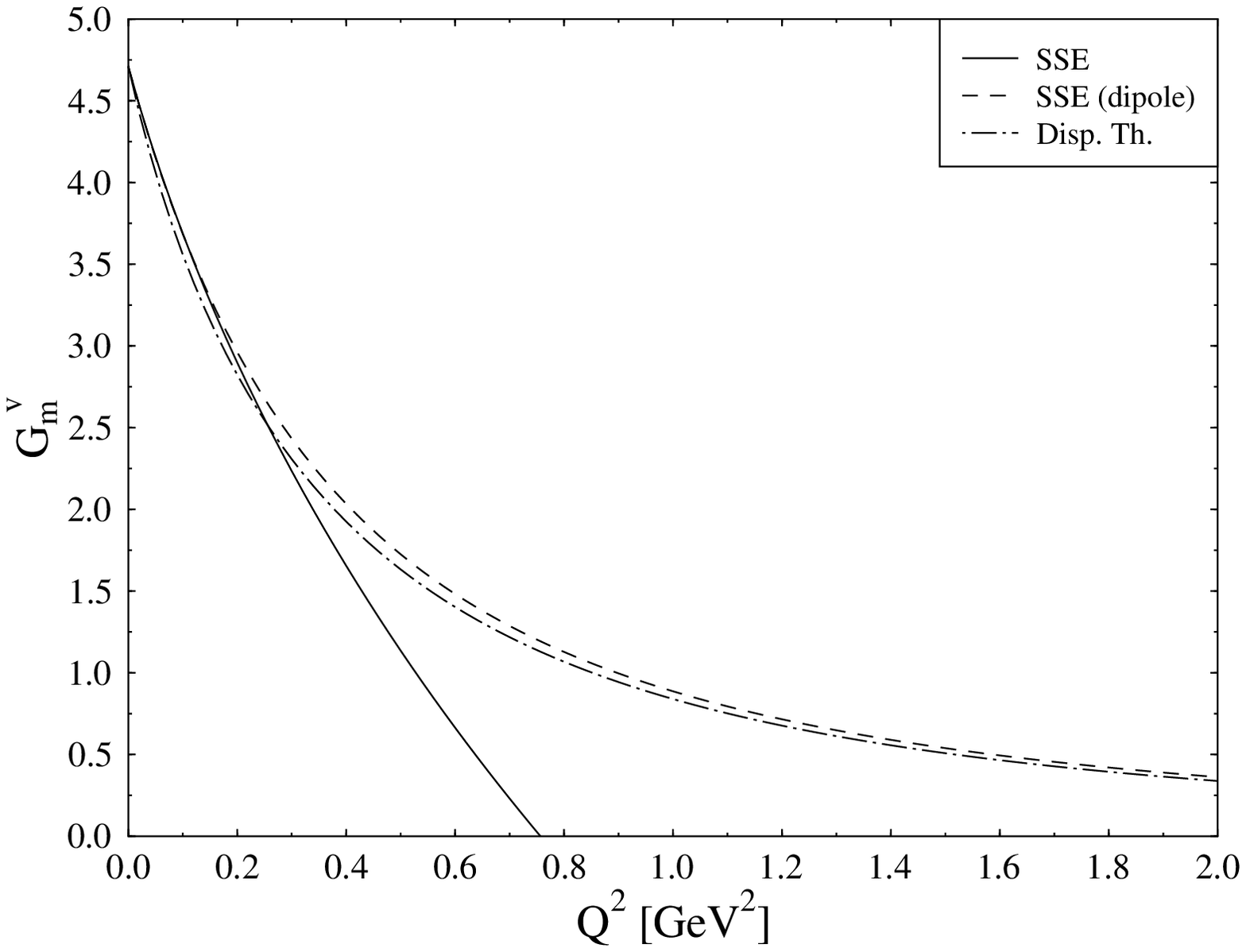} 
    \caption{Comparison of the dispersion-theoretical description
             of the isovector nucleon form factors with the SSE
             curves and the dipole approximations following from the SSE.}
    \label{fig:comp}
\end{figure}

The only difference of the above results for the form factors 
compared to the formulae given
in Ref.~\cite{chpt} lies in the mass dependence of $\kappa_v (m_\pi)$,
as the two additional diagrams (l) and (k) of Fig.~\ref{fig:diag}
do not modify the momentum dependence at $\mathcal O (\epsilon^3)$. 
The authors of Ref.~\cite{chpt} were only interested in the physical 
point $m_\pi = m_\pi^{\mathrm {phys}}$. Hence they fixed 
$\kappa_v (m_\pi^{\mathrm {phys}})$ to the empirical value 
$\kappa_v^{\mathrm {phys}} = 3.71$. In addition, one may determine
the counterterm $B_{10}^{(r)}$ such that the phenomenological
value of the isovector Dirac radius $r_1^v$ is reproduced. This leads to
$B_{10}^{(r)}(600 \, \text{MeV}) = 0.34$. 
Using for the other parameters
the phenomenological values as given in Table~\ref{tab:chipar}
and inserting (\ref{F1}) and (\ref{F2}) in (\ref{sachs})
one gets a rather good agreement with the 
experimental Sachs form factors for values of $Q^2$ up to 
about $0.4 \, \text{GeV}^2$, as exemplified in Fig.~\ref{fig:comp} by a 
comparison with the dispersion-theoretical description~\cite{mergell} 
of the isovector form factors. In addition we show in Fig.~\ref{fig:comp}
the dipole approximations derived from the SSE formulae, which will be 
explained in Sec.~\ref{sect:chpt.rad}.

\begin{table}
\caption{Empirical values of the parameters.}
\label{tab:chipar}
\begin{ruledtabular}
\begin{tabular}{cd}
Parameter   & \multicolumn{1}{c}{Empirical value} \\
\hline
$g_A$       &   1.267        \\
$c_A$       &   1.125        \\
$F_\pi$     &   0.0924 \text{ GeV}   \\
$M_N$       &   0.9389 \text{ GeV}   \\
$\Delta$    &   0.2711 \text{ GeV}   \\
$\kappa_v^{\mathrm {phys}}$  &   3.71   \\      
$\kappa_s^{\mathrm {phys}}$  &   -0.12        
\end{tabular}
\end{ruledtabular}
\end{table}

Here we want to study the quark-mass dependence of the form factors. 
Strictly speaking, in such a study all the parameters should be 
taken in the chiral limit. To order $\epsilon^3$ in the SSE,
the $m_\pi$ dependence of $F_1$ and $F_2$ is then given by the 
expressions (\ref{F1}), (\ref{F2}), (\ref{kappa}). For comparison we 
note that in Ref.~\cite{chpt} a function $\kappa_v(m_\pi)$ was 
found which corresponds to scheme $B$ in the language
of Ref.~\cite{chiralmag}. In this latter paper, scheme $B$ was however 
shown to be insufficient to describe large-mass lattice data while 
scheme $C$ turned out to work much better.
Another recent calculation \cite{kubis} of the nucleon 
form factors utilizes a relativistic framework for baryon 
chiral perturbation theory. However, as demonstrated in 
Ref.~\cite{chiralmag}, it is not able to describe the mass dependence 
of the lattice data for $\kappa_v$.
Therefore we shall not consider it for our fits.

Unfortunately, for most of the parameters the values
in the chiral limit are only poorly known. That is why we shall 
usually work with the phenomenological numbers as given in 
Table~\ref{tab:chipar} with the notable exception of the anomalous
magnetic moment.

\subsection{Form factor radii}
\label{sect:chpt.rad}

From our lattice simulations we only have data for 
values of $Q^2$ which barely touch the interval 
$0 < Q^2 < 0.4 \, \text{GeV}^2$. Therefore a direct comparison with 
(\ref{F1}) and (\ref{F2}) does not make sense (although the $Q^2$ range
in which the leading one-loop results of Eqs.~(\ref{F1}) and (\ref{F2}) 
are applicable could depend on $m_\pi$)
and we have to resort to another procedure, which exploits the 
dipole fits of our lattice form factors (see Sec.~\ref{sect:disc}).

The dipole masses of the form factors are closely related to 
the radii $r_i$ defined by the Taylor expansion of $F_i$
around $q^2=0$:
\begin{equation}  
F_i (q^2)=F_i (0) \left[ 1 + \frac{1}{6} \; r_i^2 q^2
                               + {\mathcal O}(q^4) \right] \,.
\end{equation}
If one describes the Sachs form factors by the dipole formulae 
\begin{eqnarray}  
G_e (q^2) & = & \frac{1}{(1 + Q^2/M_e^2)^2} \,, 
\nonumber \\
G_m (q^2) & = & \frac{G_m(0)}{(1 + Q^2/M_m^2)^2} \,,
\end{eqnarray}
the masses $M_e$ and $M_m$ are related to the above radii by 
\begin{equation}  \label{dipm}
\frac{1}{M_e^2} = \frac{r_1^2}{12} + \frac{\kappa}{8 M_N^2} \quad , \quad
\frac{1}{M_m^2} = \frac{r_1^2 + \kappa \, r_2^2}{12(1 + \kappa)} \,.
\end{equation}
We note again that we do not demand the two dipole masses to be equal. 
Hence violations of the uniform dipole behavior can be accounted for.

From Eqs.~(\ref{F1}) and (\ref{F2}) we can calculate the radii 
to ${\mathcal O}(\epsilon^3)$ in SSE. For the isovector 
Dirac radius one obtains
\begin{multline} \label{r1}
 \left(r_{1}^{v}\right)^2 =
    -  \frac{1}{(4\pi F_\pi)^2}\left\{1+7 g_{A}^2 +
  \left(10 g_{A}^2 +2\right) \log\left[\frac{m_\pi}{\lambda}\right]\right\} -
    \frac{12 B_{10}^{(r)}(\lambda)}{(4\pi F_\pi)^2} 
\\ 
     +  \frac{c_A^2}{54\pi^2 F_{\pi}^2}\Bigg\{
        26+30\log\left[\frac{m_\pi}{\lambda}\right]
          +30\frac{\Delta}{\sqrt{\Delta^2-m_{\pi}^2}}
             \log\left[\frac{\Delta}{m_\pi}
            +\sqrt{\frac{\Delta^2}{m_{\pi}^2}-1}\right] \Bigg\} \,. 
\end{multline}
The terms in the first bracket of Eq.~(\ref{r1}) originate from 
Goldstone boson dynamics around a spin 1/2 nucleon (diagrams (a)-(f) 
in Fig.~\ref{fig:diag}), the counter term $B_{10}^{(r)}(\lambda)$,
which depends on the regularization scale $\lambda$,
parametrizes short-distance contributions to the
Dirac radius (``the nucleon core''), and the terms in the second
bracket arise from Goldstone boson fluctuations around an intermediate 
$\Delta$(1232) state (diagrams (g)-(l) in Fig.~\ref{fig:diag}). 
Evaluating these terms at an intermediate regularization scale 
$\lambda=600 \, \text{MeV}$ with the parameters given in 
Table~\ref{tab:chipar} one obtains
\begin{equation}
\label{r1num}
(r_1^v)^2 = \left(0.41\,(N\pi)+0.29\,(\Delta\pi)\right)\,\text{fm}^2
- \frac{12 \, B_{10}^{(r)}(600 \, \text{MeV})}{(4 \pi F_\pi)^2} \,.
\end{equation} 
Note that the total result for $(r_1^v)^2$ depends only rather
weakly on the regularization scale when 
$\lambda$ varies between 500 and 700 MeV, as the scale dependence of the
$N\pi$ and the $\Delta\pi$ contributions works in opposite direction.

Compared to the empirical value 
$(r_1^v)^2_{\mathrm {exp}} = 0.585 \, \text{fm}^2$~\cite{mergell}
the leading one-loop contributions from the Goldstone
boson cloud tend to overestimate the Dirac radius (squared) by
20\%. In Ref.~\cite{chpt} it was argued that one can always adjust the
short-distance counter term $B_{10}^{(r)}(\lambda)$ 
to reproduce the physical isovector Dirac radius, e.g., 
$B_{10}^{(r)}(600 \, \text{MeV}) = 0.34$ works for the parameters of 
Table~\ref{tab:chipar}.

Here, however, we do not want to follow this philosophy. It would
mean that the leading contribution of the ``nucleon core'' to the
square of the isovector Dirac radius becomes {\em negative}.
We consider such a scenario as unphysical. 
In the following we therefore set $B_{10}^{(r)}(600 \, \text{MeV}) = 0$
(vanishing core contribution) and conclude that the
${\mathcal O}(\epsilon^3)$ SSE formula of Eq.~(\ref{r1}) is not accurate enough
to describe the quark-mass dependence of the isovector Dirac radius. 
Hence we can only expect a qualitative picture of the chiral
extrapolation curve for this quantity, as shown in 
Sec.~\ref{sect:disc.combi}. 

For the leading one-loop isovector Pauli radius (squared) one obtains
\begin{equation} \label{r2}
\left(r_{2}^{v}\right)^2 = 
   \frac{g_{A}^2 M_N}{8 F_{\pi}^2 \kappa_v (m_\pi) \pi m_\pi}
  +\frac{c_A^2 M_N}{9 F_{\pi}^2 \kappa_v (m_\pi)
                  \pi^2 \sqrt{\Delta^2-m_{\pi}^2}} 
    \log\left[\frac{\Delta}{m_\pi}+\sqrt{\frac{\Delta^2}{m_{\pi}^2}-1}\right] 
  + \frac{24 M_N}{\kappa_v(m_\pi)}\,B_{c2} \,.  
\end{equation}
The leading non-analytic quark-mass dependence $\sim m_\pi^{-1}$ is
generated via the Goldstone boson cloud around a nucleon (diagrams (a)-(f) 
of Fig.~\ref{fig:diag}), whereas the corresponding diagrams with an 
intermediate $\Delta$(1232) state (diagrams (g)-(k) in Fig.~\ref{fig:diag}) 
produce the remaining quark-mass dependence.

At leading one-loop order, in standard chiral
counting one would not encounter the term $\propto B_{c2}$ 
(see Eq.~(\ref{F2})) which parametrizes the short-distance
(``core'') contributions to the Pauli radius analogous to
$B_{10}^{(r)}(\lambda)$ in the Dirac radius~(\ref{r1}). 
However, such a term -- which should be present according to the 
physics reasoning alluded to above -- is known to exist, 
see term no.\ 54 in Ref.~\cite{fettes}.
Utilizing the parameters of Table~\ref{tab:chipar} one finds 
(for the physical pion mass) the following contributions to the radius:
\begin{equation}
(r_2^v)^2 = \left(0.53\,(N\pi)+0.09\,(\Delta\pi)
 +0.24\,\text{GeV}^3\,B_{c2}\right)\text{fm}^2\,,
\end{equation}
which without the ``core term'' $\propto B_{c2}$
are too small by 20\% when compared to the empirical value
$(r_2^v)^2_{\mathrm {exp}}=0.797 \, \text{fm}^2$ \cite{mergell}. 
Setting $B_{c2}=0.74 \, \text{GeV}^{-3}$ for the physical 
parameters considered here (see Table~\ref{tab:chipar})
one can reproduce the dispersion theoretical result with a {\em 
positive} core contribution. We shall study the chiral extrapolation function 
of $(r_2^v)^2$ with and without this core term 
in Sec.~\ref{sect:disc.combi} to test
whether our physical intuition regarding this structure holds true. 

The radii (\ref{r1}) and (\ref{r2}) display much fewer quark-mass 
dependent terms than
$\kappa_v(m_\pi)$ in Eq.~(\ref{kappa}) though all three quantities are
calculated to the same ${\mathcal O}(\epsilon^3)$ accuracy in SSE. 
This seems to have its origin in the fact that one has to take out 
a factor of $q^2$ from the ${\mathcal O}(\epsilon^3)$ expression for 
the form factors in Eqs.~(\ref{F1}), (\ref{F2})
in order to obtain the radius, leaving only a few possible structures
for quark-mass dependent terms at this order.
From the point of view of ChEFT it is therefore more involved 
to get the quark-mass dependence of radii under control than it is 
to study the quark-mass dependence of the form factors at $q^2=0$. 
In Sec.~\ref{sect:disc} we shall compare the ChEFT formulae with 
the lattice data.

Even without the additional core term in Eq.~(\ref{r2})
the dipole formulae with the above expressions for the
radii reproduce the dispersion-theoretical form factors quite 
accurately for small and moderate values of $Q^2$ as can be seen 
from the dashed  curves in Fig.~\ref{fig:comp}. This observation 
constitutes a further argument in favor of our dipole fits.
Empirical isovector dipole masses can be computed from the phenomenological
isovector radii. One finds $M_e^v = 0.75 \, \text{GeV}$ and 
$M_m^v = 0.79 \, \text{GeV}$.

\begin{table*}
\caption{Fit values from fits of Eqs.~(\ref{kappa}) and (\ref{kappas})
         to lattice data.}
\label{tab:chimag}
\begin{ruledtabular}
\begin{tabular}{cd}
Parameter   & \multicolumn{1}{c}{Value from Ref.~\cite{chiralmag}} \\ 
\hline
$\kappa_v^0$   &   5.1(4)  \\
$c_V$       &    -2.26(6) \, \text{GeV}^{-1} \\
$E_{1}^{(r)}(0.6 \,\text{GeV})$ &   -4.93(10) \, \text{GeV}^{-3} \\  
$\kappa_s^0$   &   -0.11   \\
$E_2$          &   0.074 \, \text{GeV}^{-3} \\
\end{tabular}
\end{ruledtabular}
\end{table*}

A final remark concerns the applicability of the above formulae to
quenched data. Obviously, standard ChEFT
presupposes the presence of sea quarks. However, as first unquenched
simulations show, there is little difference between quenched and unquenched 
results at presently accessible quark masses. It is therefore not unreasonable
to compare (\ref{F1}), (\ref{F2}) and (\ref{kappa}) with quenched data. 
Alternatively, one could try to develop quenched chiral perturbation
theory for the form factors. For first attempts see, e.g., 
Refs.~\cite{wilcox,savage}.
On the other hand, the size of our quark masses may lead to doubts
on the applicability of one-loop ChEFT results. Only further investigations
can clarify this issue. Here we simply try to find out how far we can
get with the available formulae.

\section{Comparison with chiral effective field theory}
\label{sect:disc}

\subsection{Comparison with previous extrapolations for $\kappa_v (m_\pi)$}
\label{sect:disc.comp}

\begin{figure}
  \includegraphics[width=12.0cm]{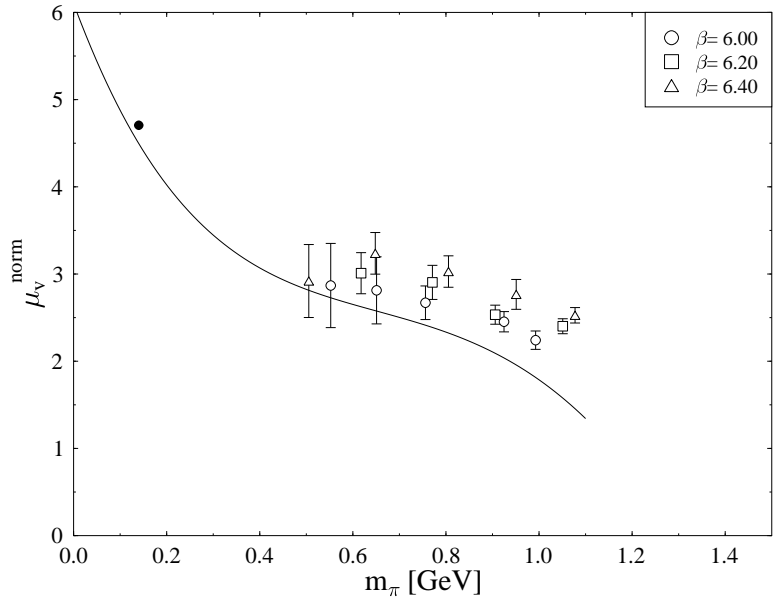}
    \caption{Our results for the isovector (normalized) magnetic moments 
             compared with the SSE extrapolation curve of 
             Ref.~\cite{chiralmag}. The solid circle represents the 
             experimental value of $\mu_v$.}
    \label{fig:magmom}
\end{figure}

Hemmert and Weise~\cite{chiralmag} fitted lattice results for 
the normalized isovector magnetic moment 
$\mu_v^{\mathrm {norm}}$ with the ${\mathcal O}(\epsilon^3)$ 
formula (\ref{kappa}) using $\kappa_v^0$, $c_V$ and $E_1^{(r)} (\lambda)$
as fit parameters and fixing the other parameters at their phenomenological
values (see Table~\ref{tab:chipar}). 
Their fit yielded a rather strong $m_\pi$ dependence of 
$\mu_v^{\mathrm {norm}}$ for small $m_\pi$. 
The values they obtained for their fit parameters are given in 
Table~\ref{tab:chimag}.
A similarly strong $m_\pi$ dependence had already
been observed in Refs.~\cite{australia1a,australia1b} for the 
magnetic moments of the proton and the neutron.
In Fig.~\ref{fig:magmom} we plot our data together with the curve
corresponding to the fit of Ref.~\cite{chiralmag}. The comparison
indicates that the data used in Ref.~\cite{chiralmag} lie somewhat 
below ours.

\subsection{Combined fits}
\label{sect:disc.combi}

The results of Ref.~\cite{chiralmag} show that using the SSE it is possible
to connect the experimental value of the magnetic moment with the lattice 
data. This raises the question
whether one could not obtain a similarly good description of the 
radii by fitting the SSE expression to the simulation results.
From the point of view of ChEFT the mass dependence of the Dirac and 
Pauli radius is much simpler to discuss than that of the analogous
Sachs quantities. Hence we shall base our analysis on $r_1^v$ and $r_2^v$
instead of $M_e^v$ and $M_m^v$.
Note, however, that the numerical data in the following discussion are 
taken from the dipole fits of the Sachs form factors.

\begin{figure}
  \includegraphics[width=12.0cm]{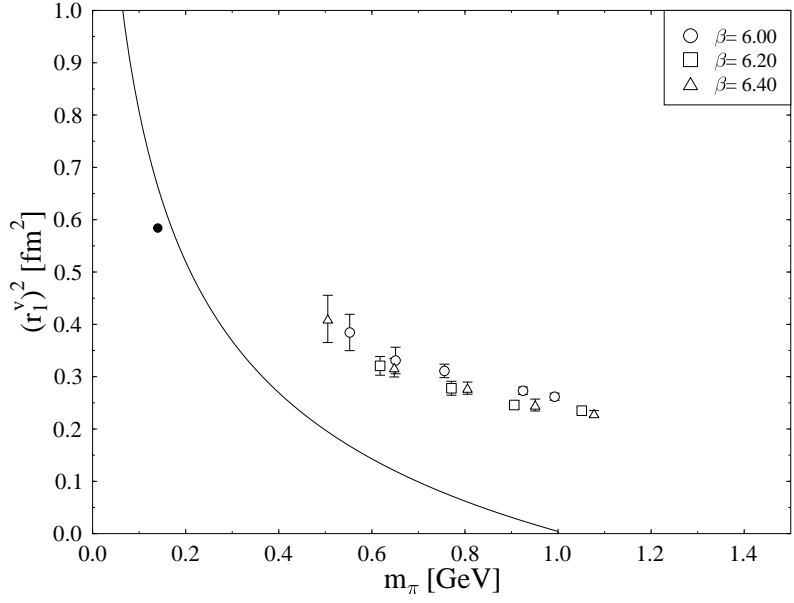} \\
  \includegraphics[width=12.0cm]{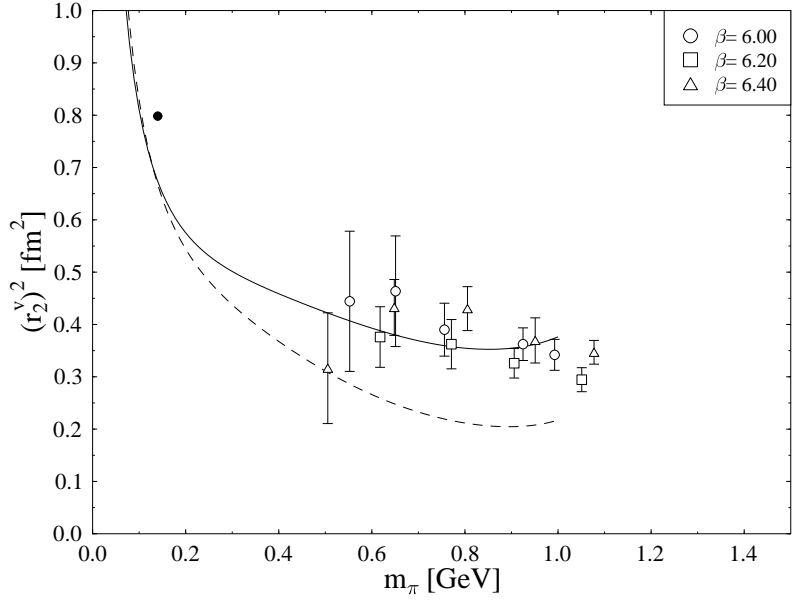} 
    \caption{Isovector radii compared with fit curves. For the fit 
             parameters see Table~\ref{tab:combifit}.
             The dashed line corresponds to the fit without core term.
             The solid circles represent the experimental values.}
    \label{fig:dipmassfit}
\end{figure}

\begin{figure}
  \includegraphics[width=12.0cm]{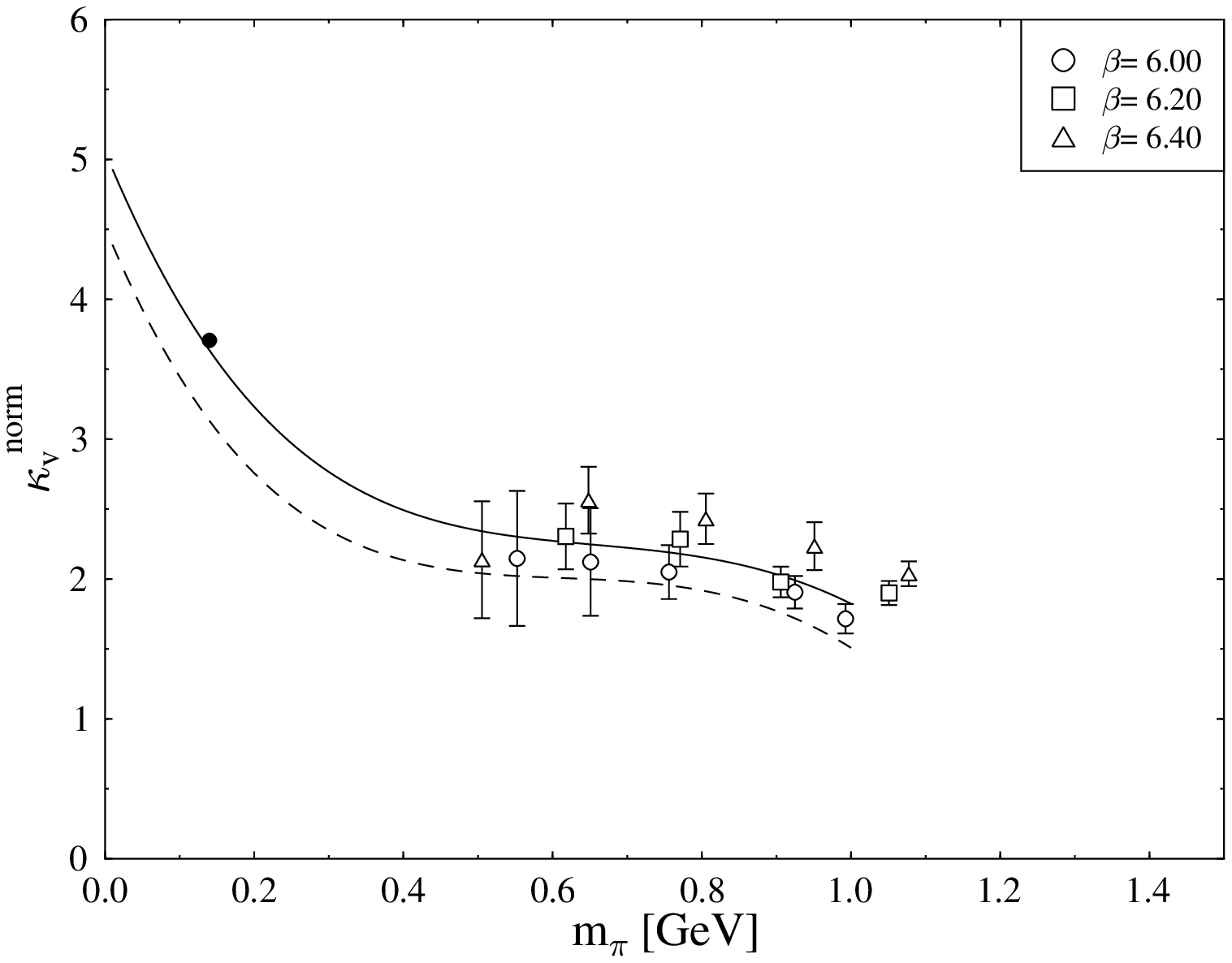}
    \caption{Isovector (normalized) anomalous magnetic moments 
             compared with (combined) fit.  For the fit parameters see 
             Table~\ref{tab:combifit}. The dashed line corresponds 
             to the fit without core term. The solid circle 
             represents the experimental value of $\kappa_v$.}
    \label{fig:magmomfit}
\end{figure}

Because cut-off effects seem to be small we fitted the results 
from all three $\beta$ values together taking into account all 
data points with $m_\pi < 1 \, \text{GeV}$. We kept $F_\pi$, $M_N$, $c_A$ 
and $\Delta$ at their phenomenological values (see Table~\ref{tab:chipar}), 
fixed the renormalization scale $\lambda$
at $0.6 \, \text{GeV}$ and chose $B_{10}^{(r)}(0.6 \, \text{GeV}) = 0$
for the reason explained in Sec.~\ref{sect:chpt.rad}. Furthermore, we set 
$g_A = 1.2$, which is the value in the chiral limit obtained in a 
recent ChEFT analysis~\cite{ga} of quenched lattice data.
This leaves us with four fit parameters: $\kappa_v^0$, $c_V$, 
$E_{1}^{(r)}(0.6 \,\text{GeV})$ and $B_{c2}$. 
As the Dirac radius $r_1^v$
is independent of these parameters, we performed a simultaneous fit
of $(r_2^v)^2(m_\pi)$ and $\kappa_v^{\mathrm {norm}}(m_\pi)$. 
The results are collected in the second column of 
Table~\ref{tab:combifit}.
Plots of our data together with the fit curves are 
shown in Figs.~\ref{fig:dipmassfit}, \ref{fig:magmomfit}.
Leaving out the core term in $r_2^v$, i.e.\ setting $B_{c2}=0$, leads to the 
parameter values given in the third column of 
Table~\ref{tab:combifit}.
The corresponding curves are shown as dashed lines 
in the figures.

\begin{table*}
\caption{Results of a combined fit (with and without core term)
         of isovector Pauli radii and anomalous magnetic moments.}
\label{tab:combifit}
\begin{ruledtabular}
\begin{tabular}{cdd}
Parameter   & \multicolumn{1}{c}{Fitted value} & 
                               \multicolumn{1}{c}{Fitted value} \\
 {}    & {}       & \multicolumn{1}{c}{without core term} \\
\hline
$\kappa_v^0$ & 5.1(8) &  4.5(9) \\
$c_V$        &  -2.3(5) \, \text{GeV}^{-1} 
             &  -2.5(6) \, \text{GeV}^{-1} \\
$E_{1}^{(r)}(0.6 \,\text{GeV})$ 
             &   -4.8(8) \, \text{GeV}^{-3} 
             &   -5.1(9) \, \text{GeV}^{-3} \\
$B_{c2}$ &  0.41(4) \, \text{GeV}^{-3} 
         &  0.0 \, \text{GeV}^{-3} \\
$\chi^2$ &  19.2 &   185.9 
\end{tabular}
\end{ruledtabular}
\end{table*}

The lattice data for the isovector anomalous magnetic moment
are very well described by the chiral extrapolation curve, in 
particular if one allows for a (small) core contribution via $B_{c2}$. 
Interestingly, the chiral extrapolation function comes rather 
close to the physical point, although the lightest 
lattice points are quite far from the physical world and large curvature 
is required. Moreover, the chiral limit value $\kappa_v^0$
and the values of the other two fit parameters $E_1^{(r)}$ and $c_V$ 
in the second column of Table~\ref{tab:combifit}
compare astonishingly well with the numbers found in Ref.~\cite{chiralmag}
(see Table~\ref{tab:chimag})
providing us with some confidence in their determination. The lattice
data for the isovector Pauli radius (squared) are also reasonably well 
described by the chiral extrapolation function of Eq.~(\ref{r2}), 
at least for pion masses below 800 MeV. The effect of 
the finite core size of the nucleon (parametrized via $B_{c2}$) is 
more visible in this quantity than in $\kappa_v$. 
While the phenomenological value at the physical point is missed by our 
central curve for $r_2^v(m_\pi)$, it would lie within
the error band, given the relatively large errors of the fit parameters.
We note that the $1/m_\pi$ chiral singularity shows up rather strongly,
dominating the curvature out to pion masses around $0.3 \, \text{GeV}$.

While our generalization of the ChEFT analysis of Ref.~\cite{chiralmag}
describes the ``magnetic'' quantities $\kappa_v$ and $r_2^v$ reasonably
well, it is not successful for the isovector Dirac radius. 
As can be seen in Fig.~\ref{fig:dipmassfit}, the chiral extrapolation 
function drops too fast with $m_\pi$ and even reaches zero around 
$m_\pi=1 \, \text{GeV}$. Remember that Dirac radius data were not
included in the fit and the curve shown corresponds to the 
``no-core term'' scenario with $B_{10}^{(r)}(\lambda=0.6 \, \text{GeV})=0$. 
One could improve the agreement between the lattice data 
and the chiral extrapolation curve by allowing $B_{10}^{(r)}$ to provide 
a positive core contribution, which would shift the 
curve upwards towards the data. However, this would result in 
extremely large values for $(r_1^v)^2$ at the physical point, 
as the shape is not modified by $B_{10}^{(r)}$. On the other 
hand, the simulation data themselves look completely reasonable,
indicating that for pion masses around $1 \, \text{GeV}$, for which
the pion cloud should be considerably reduced, the square of the Dirac
radius of the nucleon has shrunk to $\approx 0.25 \, \text{fm}^2$,
less than half of the value at the physical point. One reason for
this failure of Eq.~(\ref{r1}) might lie in important higher 
order corrections in ChEFT which could soften the strong $m_\pi$ 
dependence originating from the chiral logarithm.

Nevertheless, one should also not forget that here we are dealing 
with a quenched simulation. 
Given that $(r_1^v)^2$ at the physical point is nearly completely 
dominated by the pion cloud (for low values of $\lambda$, 
cf.\ Eq.~(\ref{r1num})) it is conceivable that the Dirac radius of the 
nucleon might be sensitive to the effects 
of (un)quenching. We therefore conclude that especially for 
$r_1^v$ a lot of work remains to be done, both on the level of 
ChEFT, where the next-to-leading one-loop contributions need to 
be evaluated, as well as on the level of the simulations, where a 
similar analysis as the one presented here has to be performed 
based on fully dynamical configurations. 

Of course, one can think of alternative fit strategies, which differ
by the choice of the fixed parameters. For example, one might leave also 
$c_A$ and $\Delta$ free in addition to the four parameters used above.
In this (or a similar) way it is possible to force the fit through the
data points for $(r_1^v)^2$ also, but then the physical point is missed
by a considerable amount. So we must conclude that at the present level
of accuracy the SSE expression for the Dirac radius is
not sufficient to connect the Monte Carlo data
in a physically sensible way with the phenomenological value.

\subsection{Beyond the isovector channel}
\label{sect:disc.iso}

While ChEFT (to the order considered in Ref.~\cite{chiralmag}) yields
the rather intricate expression (\ref{kappa}) for the 
quark-mass dependence of the isovector 
anomalous magnetic moment, the analogous expression for 
the isoscalar anomalous magnetic moment $\kappa_s = \kappa_p + \kappa_n$
of the nucleon is much simpler, 
\begin{equation} \label{kappas}
\kappa_s (m_\pi) = \kappa_s^0 - 8 E_2 M_N m_\pi^2 \,,
\end{equation}
because the Goldstone boson contributions to this quantity only
start to appear at the two-loop level~\cite{bkm}.
The new counterterm $E_2$ parametrizes quark-mass dependent 
short-distance contributions to $\kappa_s$. The error bars of the 
lattice data are quite large compared to the
small isoscalar anomalous magnetic moment. Therefore, any 
analysis based on (\ref{kappas}) and the present lattice results must be 
considered with great caution, the more so, since the lattice data are
also afflicted with the problem of the disconnected contributions. In spite
of all these caveats, we now turn to a discussion of the magnetic moments
and combinations of them which are not purely isovector quantities.

\begin{figure}
  \includegraphics[width=12.0cm]{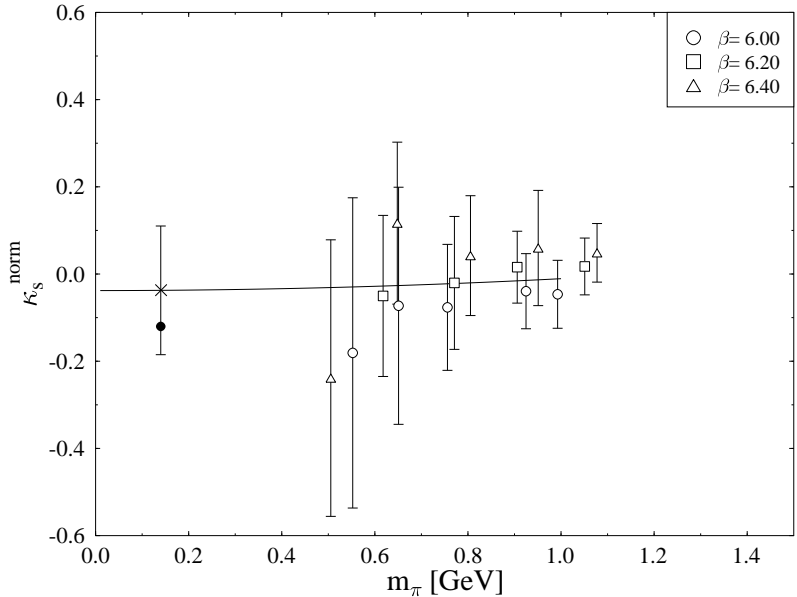}
    \caption{Isoscalar (normalized) anomalous magnetic moments 
             compared with SSE fit. The solid circle represents the 
             experimental value of $\kappa_s$. The cross with the 
             attached error bar shows the value at $m_\pi^{\mathrm {phys}}$.}
    \label{fig:kappas}
\end{figure}

In Fig.~\ref{fig:kappas} we present the normalized values of $\kappa_s$
together with a fit using Eq.~(\ref{kappas}). 
The values of $\kappa_s$ have been computed as $\kappa_p + \kappa_n$ 
from the proton and neutron dipole fits of $G_m$, and the errors 
have been determined by error propagation. We obtain 
$\kappa_s^0 = -0.04(15)$ and $E_2 = -0.004(25) \, \text{GeV}^{-3}$. 
These numbers are to be compared with the fit parameters from 
Ref.~\cite{chiralmag} given in Table~\ref{tab:chimag}.
The large statistical errors make definite statements difficult.

\begin{figure}
  \includegraphics[width=12.0cm]{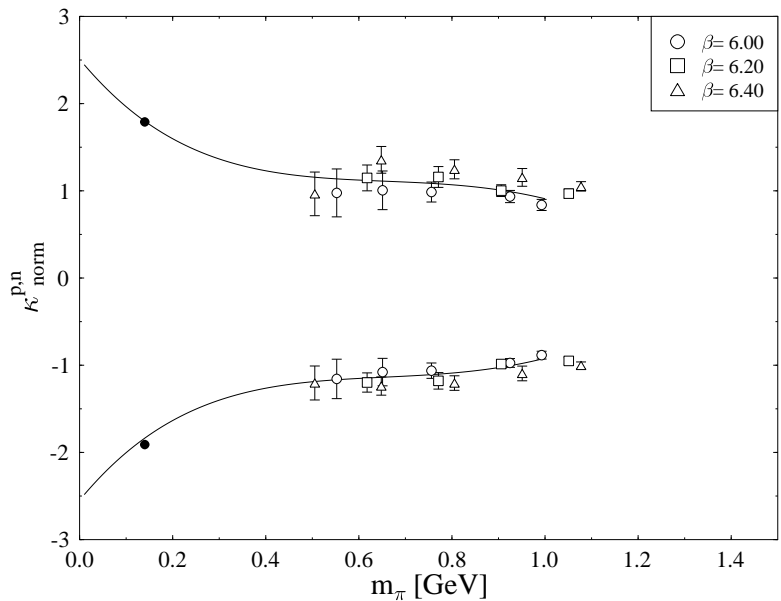}
    \caption{Anomalous magnetic moments of proton and neutron (normalized)
             with chiral extrapolation curves. The solid circles 
             represent the experimental values.}
    \label{fig:mupn}
\end{figure}

Having determined $\kappa_v(m_\pi)$ as well as $\kappa_s(m_\pi)$ 
we can now discuss the chiral extrapolation of
proton and neutron data separately.
For $\kappa_v^0$, $c_V$, $E_1^{(r)}$, $B_{c2}$ we choose the values
given in the second column of Table~\ref{tab:combifit} together with
$g_A = 1.2$, while for $\kappa_s^0$ and $E_2$ we take the numbers given 
above and the remaining parameters are fixed at their
physical values (see Table~\ref{tab:chipar}).

In Fig.~\ref{fig:mupn} we compare the resulting extrapolation functions
with the lattice results for the anomalous magnetic moments. The
extrapolation functions are surprisingly well behaved. Despite the
large gap between $m_\pi^{\mathrm {phys}}$ and the lowest data point
and the substantial curvature involved they extrapolate to the physical 
point and to the chiral limit in a very sensible way.

Finally, we want to compare our results with the predictions 
from the constituent quark model. Such comparisons are usually 
performed for ratios of observables to avoid normalization problems. 
Under the assumption
that the constituent quark mass $m_q = m_u \approx m_d$ is equal to 
$M_N/3$ also for varying $m_q$, one obtains the well-known SU(6) result
\begin{equation} 
\frac{\mu^p}{\mu^n} = 
  \frac{\mu^p_{\mathrm {norm}}}{\mu^n_{\mathrm {norm}}}
  = - \frac{3}{2}
\end{equation}
and similarly
\begin{equation} 
\frac{\kappa_p}{\kappa_n} = 
  \frac{\kappa_p^{\mathrm {norm}}}{\kappa_n^{\mathrm {norm}}}
  = - 1 \,.
\end{equation}

\begin{figure}
  \includegraphics[width=12.0cm]{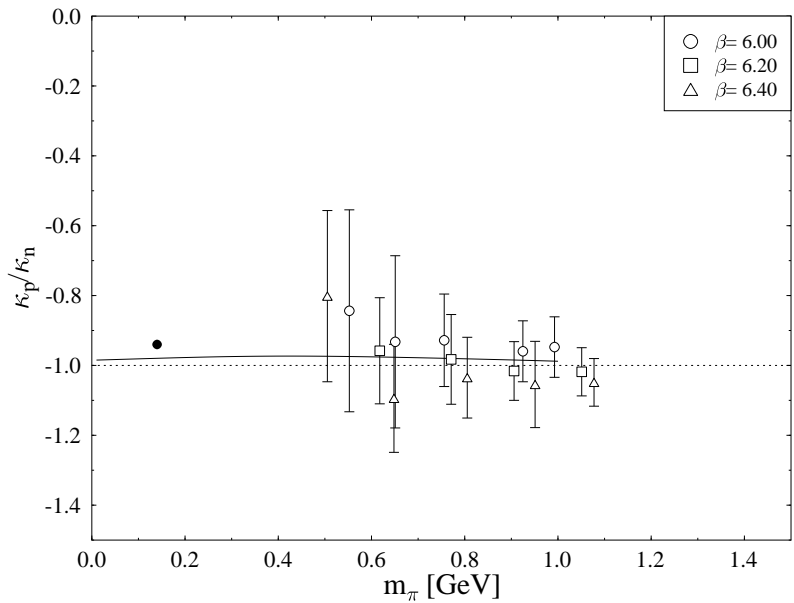}
    \caption{The ratio $\kappa_p/\kappa_n$ (identical to the ratio of 
             the normalized anomalous magnetic moments)
             with chiral extrapolation curve. The dotted line 
             shows the value $-1.0$ predicted by the non-relativistic 
             quark model. The solid circle represents the experimental value.}
    \label{fig:ratiokappab}
\end{figure}

In Fig.~\ref{fig:ratiokappab} we show the ratio of the anomalous 
magnetic moments $\kappa_p/\kappa_n$, which is identical to the ratio 
of the normalized anomalous magnetic moments, as a function of the pion mass.
The lattice data and our extrapolation function stay rather close to
the static SU(6) quark model value of $-1$ in the mass range considered
here.

\section{Conclusions}
\label{sect:concl}

We have performed a detailed study of the electromagnetic nucleon form
factors within quenched lattice QCD employing a fully non-perturbative
$O(a)$-improvement of the fermionic action and of the electromagnetic current.
Compared with previous studies 
\cite{liu,ji} we have accumulated much higher statistics, yet our statistical
errors appear to be rather large. While these older investigations used
one lattice spacing only, we have data at three different lattice 
spacings. So we could study the discretization errors and found them to 
be small.  

As the quark masses in our simulations are considerably larger than in
reality, we had to deal with chiral extrapolations.
The most effective way to handle this problem proceeds via a suitable
parametrization of the $Q^2$ dependence of the form factors. Indeed,
our data can be described reasonably well by dipole fits. 
Then the quark-mass dependence of the fit parameters (dipole masses, in
particular) can be studied. Assuming a linear dependence on the pion mass
one ends up remarkably close to the physical values, in spite of the fact
that the singularities arising from the Goldstone bosons of QCD must show 
up at some point invalidating such a simple picture. Nevertheless, the
difference between the electric and the magnetic dipole mass which we 
obtain at the physical pion mass is in (semi-quantitative) agreement 
with recent experimental results~\cite{JLAB1,JLAB2}.

Ideally, the chiral extrapolation should be guided by ChEFT.
However, most of the existing chiral expansions
do not take into account quenching artefacts and are therefore, strictly
speaking, not applicable to our data. But first simulations
with dynamical quarks indicate that at the quark masses considered in this
paper quenching effects are small so that quenched chiral
perturbation theory is not required. While in this respect the size of our
quark masses might be helpful, it leads on the other hand to doubts
on the applicability of ChEFT. Indeed, only a 
reorganisation of the standard chiral perturbation theory
series allowed Hemmert and Weise~\cite{chiralmag} to describe with a 
single expression the phenomenological value of the isovector 
anomalous magnetic moment of the nucleon as well as (quenched) lattice data. 
For a different approach to the same problem see 
Refs.~\cite{australia1a,australia1b,australia3}.

We have extended the analysis of the magnetic moments of the 
nucleon of Ref.~\cite{chiralmag} to the general case of 
nucleon electromagnetic form factors. Given that these calculations 
are reliable only for $Q^2 < 0.4 \, \text{GeV}^2$, no direct comparison 
with our lattice data, taken at higher values of $Q^2$, could be 
performed. Instead we have converted the dipole masses extracted from our
simulations into form factor radii, which could then be compared with 
the ChEFT formulae. Larger lattices allowing smaller values of 
$Q^2$ would be required, if one aims at a direct comparison with the 
ChEFT results for the form factors.

As low-order (one-loop) ChEFT is insufficient to {\em simultaneously} 
account for the quark-mass dependence of the nucleon mass and the 
form factors in the current matrix elements, we were
forced to ``normalize'' the magnetic moments computed on the lattice
before fitting them with the ChEFT formulae. Higher-order calculations 
in ChEFT, at least at the two-loop level, would be required to avoid 
this necessity.

In the isovector channel a combined fit of $\kappa_v(m_\pi)$ and 
the Pauli radius $r_2^v(m_\pi)$ yielded extrapolation functions 
which describe the lattice data quite well and extrapolate 
(albeit with large error bar) close to the physical point. 
For the isovector Dirac radius $r_1^v(m_\pi)$ 
no chiral extrapolation function could be obtained that is consistent 
both with the lattice data and known phenomenology at the physical point. 
Further studies are needed to resolve this discrepancy,
both in ChEFT regarding higher order corrections and on the 
simulation side investigating quenching effects.
(For an alternative view see Ref.~\cite{australia2}.)
The parameters obtained in the fits are well consistent 
with those found in Ref.~\cite{chiralmag}. In particular, 
we find $\kappa_v^0=5.1\pm 0.8$ as the chiral limit value for the
isovector anomalous magnetic moment of the nucleon. 

The isoscalar sector is plagued by large uncertainties in the lattice 
data. The chiral dynamics contributing to extrapolation functions in 
this sector seems to be dominated by analytic terms. Quantitative 
studies can only be performed once the statistics of the data is 
improved and disconnected contributions are taken into account.
The ratio $\kappa_p/\kappa_n$ could be well described
by our chiral extrapolation and was found in remarkable agreement 
with the constituent quark model.

The leading one-loop calculation in the SSE is 
found to describe the quark-mass dependence of magnetic quantities 
quite well. Unfortunately, at the moment we do not have a ChEFT with 
appropriate counting scheme that simultaneously describes the 
quark-mass dependence in all four quantities $\kappa_v$, $\kappa_s$, 
$r_1^v$, $r_2^v$ at leading one-loop order. It remains to be seen 
whether the discrepancies found in $r_1^v(m_\pi)$ can be resolved 
in a next-to-leading one-loop SSE calculation of the form factors.
The figures in this paper show that ChEFT often predicts large effects
at values of $m_\pi$ lighter than those we used in our lattice 
simulations. In order to confirm the predictions of ChEFT, and in order
to extrapolate reliably to physical quark masses, we need simulations
at much smaller values of $m_\pi$. Moreover, it would be desirable to
compute the quark-line disconnected contributions. Important progress 
is also to be expected from the ongoing simulations with dynamical fermions.

\begin{acknowledgments}
This work has been supported in part by 
the European Community's Human Potential Program under contract 
HPRN-CT-2000-00145, Hadrons/Lattice QCD, 
by the DFG (Forschergruppe Gitter-Hadronen-Ph\"anomenologie) 
and by the BMBF.
Discussions with V. Braun and W. Weise are gratefully acknowledged,
as well as the constructive remarks of the referee.
TRH thanks the Institute for Theoretical Physics of the University 
of Regensburg and DESY Zeuthen for their kind hospitality.

The numerical calculations were performed on the APE100 at NIC
(Zeuthen) as well as on the Cray T3E at ZIB (Berlin) and NIC
(J\"{u}lich). We wish to thank all institutions for their support.
\end{acknowledgments}

\appendix

\section{}

\begin{figure}[h]
  \includegraphics[width=16.0cm]{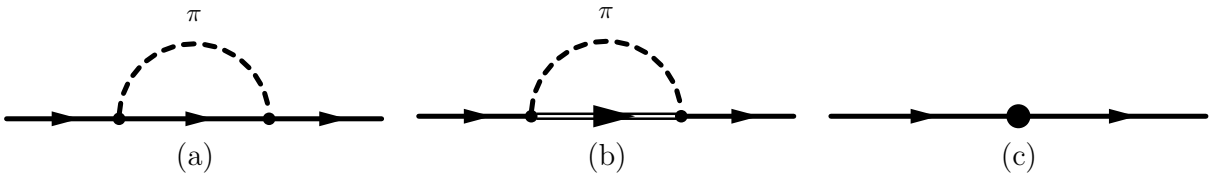}
    \caption{Diagrams in ${\mathcal O} (\epsilon^3)$ SSE contributing to the 
             nucleon mass.}
    \label{fig:diagmass}
\end{figure}

Here we want to present the nucleon mass as a function of the pion mass
in the same formalism that is used for the electromagnetic form factors,
i.e.\ in the SSE to ${\mathcal O} (\epsilon^3)$. The corresponding diagrams 
are shown in Fig.~\ref{fig:diagmass}. One finds 
\begin{multline} \label{nuclmass}
M_N = M_N^0 - 4 c_1 m_\pi^2 - \frac{3 g_A^2}{32 \pi F_\pi^2} m_\pi^3
\\
{} - \frac{c_A^2}{3 \pi^2 F_\pi^2} 
  \left \{ (\Delta^2 - m_\pi^2)^{3/2} \log R(m_\pi) 
  - (\Delta^3 - \frac{3}{2} \Delta m_\pi^2) 
                        \log \left[ \frac{2 \Delta}{m_\pi} \right]
  + \frac{1}{4} \Delta m_\pi^2 \right \} - 4 e_1 m_\pi^4 \,,
\end{multline}
where
\begin{equation}
R(m) = \frac{\Delta}{m}+\sqrt{\frac{\Delta^2}{m^2}-1} \,.
\end{equation}
In (\ref{nuclmass}) the leading correction to the nucleon mass in the 
chiral limit $M_N^0$ is parametrized by the coupling $c_1$, 
$F_\pi$ denotes the pion decay constant, $c_A$ the leading axial $N \Delta$ 
coupling, $g_A$ the axial coupling of the nucleon, $\Delta$ the 
$\Delta$(1232)-nucleon mass splitting, and $e_1$ is a counterterm. 
Eq.~(\ref{nuclmass}) generalizes the analysis of Ref.~\cite{bhm} performed
in heavy-baryon chiral perturbation theory to the SSE with dimensional 
regularization. The expected range of applicability, as reported 
in Ref.~\cite{bhm}, is therefore $m_\pi < 600 \, \text{MeV}$.

In Fig.~\ref{fig:masses} we have used $M_N^0 = 0.88 \, \text{GeV}$, 
$c_1 = - 0.93 \, \text{GeV}^{-1}$, $g_A = 1.2$, 
$e_1 = - 2.2 \, \text{GeV}^{-3}$, in accordance with phenomenological 
estimates. The remaining parameters have been fixed at their physical 
values given in Table~\ref{tab:chipar}. This choice leads to a satisfactory 
description of nucleon mass data from dynamical simulations at (relatively)
low quark masses.

\section{}

In this Appendix we collect our results for the nucleon form
factors.

\begingroup
\squeezetable
\begin{table*}[h]
\caption{Isovector nucleon form factors at $\beta = 6.0$.}
\label{tab:clover60isov}
\begin{ruledtabular}
\begin{tabular}{dddddd}
\multicolumn{1}{c}{$\kappa $} & \multicolumn{1}{c}{$a^2 Q^2$} & 
\multicolumn{1}{c}{$G_e$} & \multicolumn{1}{c}{$G_m$} & 
\multicolumn{1}{c}{$F_1$} & \multicolumn{1}{c}{$F_2$}  \\
\hline
    0.1320 &     0.0000 & 0.9962(6)         &  { }             
                        & 0.9962(6)         &  { }              \\
       { } &     0.1484 & 0.554(7)          & 2.34(4)          
                        & 0.621(7)          & 1.72(4)           \\
       { } &     0.1492 & 0.56(2)           & 2.4(2)           
                        & 0.63(2)           & 1.8(2)            \\
       { } &     0.2867 & 0.42(3)           & 1.69(17)         
                        & 0.51(3)           & 1.18(16)          \\
       { } &     0.3084 & 0.357(14)         & 1.50(8)          
                        & 0.443(14)         & 1.06(8)           \\
       { } &     0.4168 & 0.30(5)           & 1.3(2)           
                        & 0.40(5)           & 0.88(18)          \\
       { } &     0.4576 & 0.28(2)           & 1.13(8)          
                        & 0.37(2)           & 0.76(8)           \\
       { } &     0.6169 & 0.141(9)          & 0.67(4)          
                        & 0.215(9)          & 0.45(3)           \\
\hline
    0.1324 &     0.0000 & 0.9936(7)         &  { }             
                        & 0.9936(7)         &  { }              \\
       { } &     0.1480 & 0.544(7)          & 2.40(4)          
                        & 0.619(7)          & 1.78(4)           \\
       { } &     0.1488 & 0.60(2)           & 2.6(2)           
                        & 0.68(2)           & 1.88(19)          \\
       { } &     0.2852 & 0.40(3)           & 1.64(16)         
                        & 0.49(3)           & 1.15(15)          \\
       { } &     0.3084 & 0.327(14)         & 1.41(9)          
                        & 0.415(14)         & 1.00(8)           \\
       { } &     0.4137 & 0.26(3)           & 1.39(17)         
                        & 0.38(4)           & 1.01(16)          \\
       { } &     0.4573 & 0.27(2)           & 1.10(9)          
                        & 0.36(2)           & 0.73(8)           \\
       { } &     0.5350 & 0.108(18)         & 0.59(12)         
                        & 0.17(2)           & 0.42(11)          \\
       { } &     0.6169 & 0.131(8)          & 0.69(4)          
                        & 0.216(9)          & 0.48(3)           \\
\hline
    0.1333 &     0.0000 & 0.9921(18)        &  { }             
                        & 0.9921(18)        &  { }              \\
       { } &     0.1463 & 0.503(10)         & 2.21(6)          
                        & 0.590(9)          & 1.62(5)           \\
       { } &     0.1477 & 0.58(4)           & 2.6(4)           
                        & 0.68(4)           & 2.0(4)            \\
       { } &     0.2796 & 0.37(4)           & 1.5(2)           
                        & 0.48(4)           & 1.0(2)            \\
       { } &     0.3084 & 0.28(2)           & 1.21(12)         
                        & 0.38(2)           & 0.83(11)          \\
       { } &     0.4029 &  { }              & 1.3(2)           
                        &  { }              &  { }              \\
       { } &     0.4561 & 0.24(5)           & 1.02(15)         
                        & 0.35(4)           & 0.67(13)          \\
       { } &     0.6169 & 0.099(11)         & 0.59(5)          
                        & 0.191(13)         & 0.40(4)           \\
\hline
    0.1338 &     0.0000 & 0.999(4)          &  { }             
                        & 0.999(4)          &  { }              \\
       { } &     0.1447 & 0.475(14)         & 1.94(8)          
                        & 0.566(14)         & 1.38(8)           \\
       { } &     0.2741 & 0.30(6)           & 0.8(2)           
                        & 0.36(6)           & 0.5(2)            \\
       { } &     0.3084 & 0.28(3)           & 1.17(16)         
                        & 0.39(3)           & 0.78(14)          \\
       { } &     0.6169 & 0.12(2)           & 0.49(6)          
                        & 0.20(2)           & 0.29(5)           \\
\hline
    0.1342 &     0.0000 & 0.987(6)          &  { }             
                        & 0.987(6)          &  { }              \\
       { } &     0.1439 & 0.437(17)         & 1.91(9)          
                        & 0.535(17)         & 1.38(9)           \\
       { } &     0.3084 & 0.26(4)           & 0.98(17)         
                        & 0.35(4)           & 0.62(15)          \\
       { } &     0.6169 &  { }              & 0.49(8)          
                        &  { }              &  { }              \\
\end{tabular}
\end{ruledtabular}
\end{table*}
\endgroup

\begingroup
\squeezetable
\begin{table*}
\caption{Isovector nucleon form factors at $\beta = 6.2$.}
\label{tab:clover62isov}
\begin{ruledtabular}
\begin{tabular}{dddddd}
\multicolumn{1}{c}{$\kappa $} & \multicolumn{1}{c}{$a^2 Q^2$} & 
\multicolumn{1}{c}{$G_e$} & \multicolumn{1}{c}{$G_m$} & 
\multicolumn{1}{c}{$F_1$} & \multicolumn{1}{c}{$F_2$}  \\
\hline
    0.1333 &     0.0000 & 1.0010(2)         &  { }             
                        & 1.0010(2)         &  { }              \\
       { } &     0.0665 & 0.621(7)          & 3.03(5)          
                        & 0.692(7)          & 2.34(5)           \\
       { } &     0.0667 & 0.615(18)         & 2.90(17)         
                        & 0.683(18)         & 2.21(16)          \\
       { } &     0.1294 & 0.413(18)         & 2.03(11)         
                        & 0.504(18)         & 1.52(10)          \\
       { } &     0.1371 & 0.407(15)         & 2.04(10)         
                        & 0.504(15)         & 1.53(10)          \\
       { } &     0.1892 & 0.34(4)           & 1.62(17)         
                        & 0.44(4)           & 1.18(16)          \\
       { } &     0.2038 & 0.32(2)           & 1.58(8)          
                        & 0.43(2)           & 1.15(8)           \\
       { } &     0.2742 & 0.206(10)         & 1.10(5)          
                        & 0.306(11)         & 0.79(5)           \\
\hline
    0.1339 &     0.0000 & 1.0009(3)         &  { }             
                        & 1.0009(3)         &  { }              \\
       { } &     0.0661 & 0.597(7)          & 2.77(5)          
                        & 0.676(7)          & 2.10(5)           \\
       { } &     0.0664 & 0.64(2)           & 2.7(2)           
                        & 0.71(2)           & 2.0(2)            \\
       { } &     0.1279 & 0.44(3)           & 2.0(2)           
                        & 0.55(3)           & 1.50(19)          \\
       { } &     0.1371 & 0.406(16)         & 1.84(9)          
                        & 0.510(16)         & 1.34(8)           \\
       { } &     0.2035 & 0.32(2)           & 1.34(9)          
                        & 0.43(2)           & 0.92(8)           \\
       { } &     0.2742 & 0.176(11)         & 0.95(5)          
                        & 0.279(12)         & 0.67(4)           \\
\hline
    0.1344 &     0.0000 & 1.0031(7)         &  { }             
                        & 1.0031(7)         &  { }              \\
       { } &     0.0655 & 0.562(11)         & 2.74(7)          
                        & 0.658(11)         & 2.08(7)           \\
       { } &     0.0660 & 0.56(4)           & 2.8(3)           
                        & 0.66(4)           & 2.1(3)            \\
       { } &     0.1259 & 0.35(3)           & 1.64(15)         
                        & 0.46(3)           & 1.19(14)          \\
       { } &     0.1371 & 0.35(2)           & 1.65(14)         
                        & 0.46(2)           & 1.19(13)          \\
       { } &     0.2031 & 0.28(4)           & 1.43(14)         
                        & 0.43(3)           & 1.00(12)          \\
       { } &     0.2742 & 0.162(15)         & 0.87(7)          
                        & 0.277(17)         & 0.59(6)           \\
\hline
    0.1349 &     0.0000 & 1.0052(18)        &  { }             
                        & 1.0052(18)        &  { }              \\
       { } &     0.0647 & 0.525(12)         & 2.44(7)          
                        & 0.631(12)         & 1.81(7)           \\
       { } &     0.0654 & 0.55(5)           & 2.8(4)           
                        & 0.67(5)           & 2.1(4)            \\
       { } &     0.1233 & 0.30(4)           & 1.44(19)         
                        & 0.42(4)           & 1.02(18)          \\
       { } &     0.1371 & 0.31(3)           & 1.41(14)         
                        & 0.44(3)           & 0.97(13)          \\
       { } &     0.2025 & 0.26(4)           & 1.24(14)         
                        & 0.42(4)           & 0.82(13)          \\
       { } &     0.2742 & 0.123(18)         & 0.75(7)          
                        & 0.25(2)           & 0.51(6)           \\
\end{tabular}
\end{ruledtabular}
\end{table*}
\endgroup

\begingroup
\squeezetable
\begin{table*}
\caption{Isovector nucleon form factors at $\beta = 6.4$.}
\label{tab:clover64isov}
\begin{ruledtabular}
\begin{tabular}{dddddd}
\multicolumn{1}{c}{$\kappa $} & \multicolumn{1}{c}{$a^2 Q^2$} & 
\multicolumn{1}{c}{$G_e$} & \multicolumn{1}{c}{$G_m$} & 
\multicolumn{1}{c}{$F_1$} & \multicolumn{1}{c}{$F_2$}  \\
\hline
    0.1338 &     0.0000 & 1.0019(18)        &  { }             
                        & 1.0019(18)        &  { }              \\
       { } &     0.0375 & 0.636(6)          & 3.10(5)          
                        & 0.705(6)          & 2.40(5)           \\
       { } &     0.0376 & 0.626(16)         & 3.05(14)         
                        & 0.693(16)         & 2.36(14)          \\
       { } &     0.0730 & 0.413(18)         & 1.95(10)         
                        & 0.494(18)         & 1.45(10)          \\
       { } &     0.0771 & 0.416(13)         & 2.10(8)          
                        & 0.510(13)         & 1.59(7)           \\
       { } &     0.1069 & 0.29(2)           & 1.34(9)          
                        & 0.37(2)           & 0.97(9)           \\
       { } &     0.1147 & 0.30(2)           & 1.48(7)          
                        & 0.397(19)         & 1.09(7)           \\
       { } &     0.1394 & 0.21(2)           & 1.00(14)         
                        & 0.29(2)           & 0.71(13)          \\
       { } &     0.1542 & 0.215(11)         & 1.13(5)          
                        & 0.311(11)         & 0.82(5)           \\
\hline
    0.1342 &     0.0000 & 1.002(5)          &  { }             
                        & 1.002(5)          &  { }              \\
       { } &     0.0373 & 0.611(11)         & 3.01(8)          
                        & 0.689(11)         & 2.32(7)           \\
       { } &     0.0374 & 0.61(3)           & 3.0(2)           
                        & 0.69(3)           & 2.3(2)            \\
       { } &     0.0724 & 0.44(4)           & 2.2(2)           
                        & 0.55(4)           & 1.7(2)            \\
       { } &     0.0771 & 0.38(2)           & 2.05(14)         
                        & 0.49(2)           & 1.56(13)          \\
       { } &     0.1145 & 0.26(3)           & 1.38(11)         
                        & 0.36(3)           & 1.02(11)          \\
       { } &     0.1542 & 0.171(19)         & 0.93(8)          
                        & 0.26(2)           & 0.66(8)           \\
\hline
    0.1346 &     0.0000 & 1.003(5)          &  { }             
                        & 1.003(5)          &  { }              \\
       { } &     0.0370 & 0.576(10)         & 2.77(6)          
                        & 0.665(10)         & 2.11(6)           \\
       { } &     0.0372 & 0.54(3)           & 2.8(3)           
                        & 0.63(3)           & 2.1(3)            \\
       { } &     0.0713 & 0.34(3)           & 1.53(13)         
                        & 0.43(3)           & 1.10(12)          \\
       { } &     0.0771 & 0.347(19)         & 1.71(10)         
                        & 0.457(19)         & 1.25(9)           \\
       { } &     0.1034 & 0.23(4)           & 1.09(14)         
                        & 0.32(4)           & 0.76(13)          \\
       { } &     0.1143 & 0.28(4)           & 1.24(10)         
                        & 0.39(4)           & 0.85(10)          \\
       { } &     0.1338 & 0.16(4)           & 0.71(14)         
                        & 0.23(4)           & 0.48(13)          \\
       { } &     0.1542 & 0.163(14)         & 0.90(6)          
                        & 0.273(15)         & 0.63(5)           \\
\hline
    0.1350 &     0.0000 & 1.006(8)          &  { }             
                        & 1.006(8)          &  { }              \\
       { } &     0.0366 & 0.531(12)         & 2.60(8)          
                        & 0.636(12)         & 1.97(7)           \\
       { } &     0.0369 & 0.48(5)           & 2.6(4)           
                        & 0.59(6)           & 2.0(4)            \\
       { } &     0.0700 & 0.32(4)           & 1.57(19)         
                        & 0.44(4)           & 1.13(18)          \\
       { } &     0.0771 & 0.29(3)           & 1.58(15)         
                        & 0.42(3)           & 1.16(13)          \\
       { } &     0.1009 &  { }              & 1.2(3)           
                        &  { }              &  { }              \\
       { } &     0.1140 & 0.26(5)           & 1.11(12)         
                        & 0.38(5)           & 0.73(11)          \\
       { } &     0.1298 &  { }              & 0.69(17)         
                        &  { }              &  { }              \\
       { } &     0.1542 & 0.142(17)         & 0.78(6)          
                        & 0.259(18)         & 0.53(5)           \\
\hline
    0.1353 &     0.0000 & 1.006(7)          &  { }             
                        & 1.006(7)          &  { }              \\
       { } &     0.0360 & 0.48(2)           & 2.22(14)         
                        & 0.59(2)           & 1.63(13)          \\
       { } &     0.0681 &  { }              & 1.5(3)           
                        &  { }              &  { }              \\
       { } &     0.0771 & 0.24(5)           & 1.39(18)         
                        & 0.39(5)           & 1.00(17)          \\
       { } &     0.1542 &  { }              & 0.73(11)         
                        &  { }              &  { }              \\
\end{tabular}
\end{ruledtabular}
\end{table*}
\endgroup

\begingroup
\squeezetable
\begin{table*}
\caption{Proton form factors at $\beta = 6.0$.}
\label{tab:clover60prot}
\begin{ruledtabular}
\begin{tabular}{dddddd}
\multicolumn{1}{c}{$\kappa $} & \multicolumn{1}{c}{$a^2 Q^2$} & 
\multicolumn{1}{c}{$G_e$} & \multicolumn{1}{c}{$G_m$} & 
\multicolumn{1}{c}{$F_1$} & \multicolumn{1}{c}{$F_2$}  \\
\hline
    0.1320 &     0.0000 & 0.9980(5)         &  { }             
                        & 0.9980(5)         &  { }              \\
       { } &     0.1484 & 0.566(6)          & 1.43(3)          
                        & 0.599(6)          & 0.83(3)           \\
       { } &     0.1492 & 0.58(2)           & 1.49(13)         
                        & 0.61(2)           & 0.88(13)          \\
       { } &     0.2867 & 0.42(2)           & 1.02(10)         
                        & 0.47(2)           & 0.56(10)          \\
       { } &     0.3084 & 0.372(12)         & 0.91(5)          
                        & 0.413(11)         & 0.50(5)           \\
       { } &     0.4168 & 0.34(5)           & 0.77(12)         
                        & 0.38(4)           & 0.39(12)          \\
       { } &     0.4576 & 0.287(19)         & 0.70(5)          
                        & 0.332(18)         & 0.37(5)           \\
       { } &     0.6169 & 0.153(8)          & 0.41(2)          
                        & 0.190(7)          & 0.22(2)           \\
\hline
    0.1324 &     0.0000 & 0.9964(6)         &  { }             
                        & 0.9964(6)         &  { }              \\
       { } &     0.1480 & 0.557(6)          & 1.47(3)          
                        & 0.594(5)          & 0.87(3)           \\
       { } &     0.1488 & 0.600(19)         & 1.59(13)         
                        & 0.640(19)         & 0.95(13)          \\
       { } &     0.2852 & 0.40(2)           & 1.02(10)         
                        & 0.45(2)           & 0.57(9)           \\
       { } &     0.3084 & 0.340(11)         & 0.88(5)          
                        & 0.384(11)         & 0.49(5)           \\
       { } &     0.4137 & 0.29(3)           & 0.84(10)         
                        & 0.34(3)           & 0.49(10)          \\
       { } &     0.4573 & 0.267(18)         & 0.68(6)          
                        & 0.315(17)         & 0.37(5)           \\
       { } &     0.5350 & 0.118(17)         & 0.37(8)          
                        & 0.152(18)         & 0.22(7)           \\
       { } &     0.6169 & 0.143(7)          & 0.43(2)          
                        & 0.187(7)          & 0.24(2)           \\
\hline
    0.1333 &     0.0000 & 0.9957(13)        &  { }             
                        & 0.9957(13)        &  { }              \\
       { } &     0.1463 & 0.517(8)          & 1.36(3)          
                        & 0.560(8)          & 0.80(3)           \\
       { } &     0.1477 & 0.57(3)           & 1.6(2)           
                        & 0.62(3)           & 1.0(2)            \\
       { } &     0.2796 & 0.39(3)           & 0.91(14)         
                        & 0.44(3)           & 0.47(13)          \\
       { } &     0.3084 & 0.294(15)         & 0.76(7)          
                        & 0.342(15)         & 0.42(7)           \\
       { } &     0.4029 & 0.27(5)           & 0.81(15)         
                        & 0.34(5)           & 0.47(14)          \\
       { } &     0.4561 & 0.26(4)           & 0.66(9)          
                        & 0.32(3)           & 0.34(9)           \\
       { } &     0.6169 & 0.113(9)          & 0.37(3)          
                        & 0.161(9)          & 0.21(3)           \\
\hline
    0.1338 &     0.0000 & 1.000(3)          &  { }             
                        & 1.000(3)          &  { }              \\
       { } &     0.1447 & 0.488(12)         & 1.21(5)          
                        & 0.532(11)         & 0.67(5)           \\
       { } &     0.2741 & 0.30(5)           & 0.54(15)         
                        & 0.32(5)           & 0.22(14)          \\
       { } &     0.3084 & 0.30(2)           & 0.74(10)         
                        & 0.35(2)           & 0.39(9)           \\
       { } &     0.6169 & 0.125(17)         & 0.31(4)          
                        & 0.166(16)         & 0.14(3)           \\
\hline
    0.1342 &     0.0000 & 0.994(5)          &  { }             
                        & 0.994(5)          &  { }              \\
       { } &     0.1439 & 0.451(13)         & 1.17(6)          
                        & 0.499(13)         & 0.68(6)           \\
       { } &     0.3084 & 0.25(3)           & 0.62(11)         
                        & 0.30(3)           & 0.32(10)          \\
       { } &     0.6169 & 0.094(19)         & 0.31(5)          
                        & 0.145(19)         & 0.17(4)           \\
\end{tabular}
\end{ruledtabular}
\end{table*}
\endgroup

\begingroup
\squeezetable
\begin{table*}
\caption{Proton form factors at $\beta = 6.2$.}
\label{tab:clover62prot}
\begin{ruledtabular}
\begin{tabular}{dddddd}
\multicolumn{1}{c}{$\kappa $} & \multicolumn{1}{c}{$a^2 Q^2$} & 
\multicolumn{1}{c}{$G_e$} & \multicolumn{1}{c}{$G_m$} & 
\multicolumn{1}{c}{$F_1$} & \multicolumn{1}{c}{$F_2$}  \\
\hline
    0.1333 &     0.0000 & 1.00196(17)       &  { }             
                        & 1.00196(17)       &  { }              \\
       { } &     0.0665 & 0.633(5)          & 1.85(3)          
                        & 0.669(5)          & 1.18(3)           \\
       { } &     0.0667 & 0.626(15)         & 1.82(11)         
                        & 0.662(15)         & 1.16(10)          \\
       { } &     0.1294 & 0.426(15)         & 1.24(7)          
                        & 0.472(15)         & 0.77(6)           \\
       { } &     0.1371 & 0.423(12)         & 1.24(6)          
                        & 0.471(12)         & 0.77(6)           \\
       { } &     0.1892 & 0.37(4)           & 1.00(11)         
                        & 0.42(4)           & 0.58(10)          \\
       { } &     0.2038 & 0.333(18)         & 0.97(5)          
                        & 0.387(17)         & 0.58(5)           \\
       { } &     0.2742 & 0.219(10)         & 0.67(3)          
                        & 0.269(9)          & 0.40(3)           \\
\hline
    0.1339 &     0.0000 & 1.0020(3)         &  { }             
                        & 1.0020(3)         &  { }              \\
       { } &     0.0661 & 0.607(6)          & 1.70(3)          
                        & 0.646(6)          & 1.05(3)           \\
       { } &     0.0664 & 0.637(19)         & 1.71(13)         
                        & 0.676(19)         & 1.04(12)          \\
       { } &     0.1279 & 0.45(3)           & 1.22(12)         
                        & 0.50(3)           & 0.72(12)          \\
       { } &     0.1371 & 0.414(13)         & 1.13(6)          
                        & 0.466(13)         & 0.66(5)           \\
       { } &     0.2035 & 0.33(2)           & 0.83(5)          
                        & 0.380(19)         & 0.45(5)           \\
       { } &     0.2742 & 0.184(10)         & 0.58(3)          
                        & 0.237(9)          & 0.34(3)           \\
\hline
    0.1344 &     0.0000 & 1.0037(5)         &  { }             
                        & 1.0037(5)         &  { }              \\
       { } &     0.0655 & 0.576(8)          & 1.67(4)          
                        & 0.624(8)          & 1.05(4)           \\
       { } &     0.0660 & 0.56(3)           & 1.8(2)           
                        & 0.62(3)           & 1.1(2)            \\
       { } &     0.1259 & 0.37(2)           & 1.02(9)          
                        & 0.42(2)           & 0.60(9)           \\
       { } &     0.1371 & 0.368(18)         & 1.00(8)          
                        & 0.424(18)         & 0.57(8)           \\
       { } &     0.1823 & 0.29(7)           &  { }             
                        &  { }              &  { }              \\
       { } &     0.2031 & 0.30(3)           & 0.88(8)          
                        & 0.38(3)           & 0.50(8)           \\
       { } &     0.2742 & 0.174(13)         & 0.53(4)          
                        & 0.232(13)         & 0.30(4)           \\
\hline
    0.1349 &     0.0000 & 1.0054(12)        &  { }             
                        & 1.0054(12)        &  { }              \\
       { } &     0.0647 & 0.535(9)          & 1.49(5)          
                        & 0.588(9)          & 0.91(4)           \\
       { } &     0.0654 & 0.55(4)           & 1.8(3)           
                        & 0.62(4)           & 1.2(3)            \\
       { } &     0.1233 & 0.34(3)           & 0.86(12)         
                        & 0.39(3)           & 0.47(11)          \\
       { } &     0.1371 & 0.34(2)           & 0.85(8)          
                        & 0.392(19)         & 0.46(8)           \\
       { } &     0.2025 & 0.27(4)           & 0.76(9)          
                        & 0.34(3)           & 0.41(8)           \\
       { } &     0.2742 & 0.133(13)         & 0.47(4)          
                        & 0.199(13)         & 0.27(3)           \\
\end{tabular}
\end{ruledtabular}
\end{table*}
\endgroup

\begingroup
\squeezetable
\begin{table*}
\caption{Proton form factors at $\beta = 6.4$.}
\label{tab:clover64prot}
\begin{ruledtabular}
\begin{tabular}{dddddd}
\multicolumn{1}{c}{$\kappa $} & \multicolumn{1}{c}{$a^2 Q^2$} & 
\multicolumn{1}{c}{$G_e$} & \multicolumn{1}{c}{$G_m$} & 
\multicolumn{1}{c}{$F_1$} & \multicolumn{1}{c}{$F_2$}  \\
\hline
    0.1338 &     0.0000 & 1.0024(14)        &  { }             
                        & 1.0024(14)        &  { }              \\
       { } &     0.0375 & 0.644(5)          & 1.89(3)          
                        & 0.679(5)          & 1.21(3)           \\
       { } &     0.0376 & 0.641(13)         & 1.89(9)          
                        & 0.676(13)         & 1.21(9)           \\
       { } &     0.0730 & 0.421(16)         & 1.19(6)          
                        & 0.462(16)         & 0.73(6)           \\
       { } &     0.0771 & 0.430(11)         & 1.30(5)          
                        & 0.479(11)         & 0.82(5)           \\
       { } &     0.1069 & 0.30(2)           & 0.82(6)          
                        & 0.34(2)           & 0.48(6)           \\
       { } &     0.1147 & 0.311(17)         & 0.91(5)          
                        & 0.359(16)         & 0.55(4)           \\
       { } &     0.1394 & 0.22(2)           & 0.61(9)          
                        & 0.25(2)           & 0.36(8)           \\
       { } &     0.1542 & 0.225(10)         & 0.69(3)          
                        & 0.274(9)          & 0.42(3)           \\
\hline
    0.1342 &     0.0000 & 1.002(4)          &  { }             
                        & 1.002(4)          &  { }              \\
       { } &     0.0373 & 0.622(9)          & 1.84(4)          
                        & 0.661(9)          & 1.18(4)           \\
       { } &     0.0374 & 0.60(2)           & 1.86(15)         
                        & 0.65(2)           & 1.22(15)          \\
       { } &     0.0724 & 0.45(3)           & 1.38(14)         
                        & 0.50(3)           & 0.88(13)          \\
       { } &     0.0771 & 0.398(19)         & 1.26(9)          
                        & 0.454(19)         & 0.81(8)           \\
       { } &     0.1145 & 0.27(3)           & 0.84(7)          
                        & 0.32(2)           & 0.52(7)           \\
       { } &     0.1542 & 0.186(17)         & 0.58(5)          
                        & 0.234(16)         & 0.34(5)           \\
\hline
    0.1346 &     0.0000 & 1.003(3)          &  { }             
                        & 1.003(3)          &  { }              \\
       { } &     0.0370 & 0.588(8)          & 1.70(4)          
                        & 0.632(7)          & 1.06(4)           \\
       { } &     0.0372 & 0.58(2)           & 1.72(17)         
                        & 0.62(2)           & 1.10(17)          \\
       { } &     0.0713 & 0.35(2)           & 0.95(8)          
                        & 0.40(2)           & 0.55(8)           \\
       { } &     0.0771 & 0.368(15)         & 1.07(6)          
                        & 0.425(14)         & 0.65(6)           \\
       { } &     0.1034 & 0.24(4)           & 0.67(9)          
                        & 0.29(3)           & 0.38(8)           \\
       { } &     0.1143 & 0.28(3)           & 0.76(7)          
                        & 0.34(3)           & 0.42(7)           \\
       { } &     0.1338 & 0.16(2)           & 0.44(9)          
                        & 0.19(2)           & 0.24(8)           \\
       { } &     0.1542 & 0.175(12)         & 0.55(4)          
                        & 0.232(11)         & 0.32(3)           \\
\hline
    0.1350 &     0.0000 & 1.004(6)          &  { }             
                        & 1.004(6)          &  { }              \\
       { } &     0.0366 & 0.549(10)         & 1.60(5)          
                        & 0.602(9)          & 1.00(5)           \\
       { } &     0.0369 & 0.53(4)           & 1.6(3)           
                        & 0.59(4)           & 1.0(3)            \\
       { } &     0.0700 & 0.32(3)           & 1.00(12)         
                        & 0.38(3)           & 0.61(11)          \\
       { } &     0.0771 & 0.319(19)         & 0.99(9)          
                        & 0.387(19)         & 0.61(9)           \\
       { } &     0.1009 & 0.21(4)           & 0.74(18)         
                        & 0.27(4)           & 0.47(16)          \\
       { } &     0.1140 & 0.27(4)           & 0.64(8)          
                        & 0.32(4)           & 0.32(8)           \\
       { } &     0.1298 &  { }              & 0.41(10)         
                        &  { }              &  { }              \\
       { } &     0.1542 & 0.150(13)         & 0.48(4)          
                        & 0.210(13)         & 0.27(3)           \\
\hline
    0.1353 &     0.0000 & 1.002(5)          &  { }             
                        & 1.002(5)          &  { }              \\
       { } &     0.0360 & 0.474(17)         & 1.38(9)          
                        & 0.533(17)         & 0.85(8)           \\
       { } &     0.0681 & 0.32(6)           & 1.0(2)           
                        & 0.40(6)           & 0.6(2)            \\
       { } &     0.0771 & 0.28(4)           & 0.85(11)         
                        & 0.35(3)           & 0.50(10)          \\
       { } &     0.1542 & 0.15(3)           & 0.48(7)          
                        & 0.23(3)           & 0.25(6)           \\
\end{tabular}
\end{ruledtabular}
\end{table*}
\endgroup

\begin{table}
\caption{Magnetic form factor of the neutron at $\beta = 6.0$.} 
\label{tab:clover60neut}
\begin{ruledtabular}
\begin{tabular}{ddd}
\multicolumn{1}{c}{$\kappa $} & \multicolumn{1}{c}{$a^2 Q^2$} & 
\multicolumn{1}{c}{$G_m$}   \\
\hline
    0.1320 &     0.1484 & -0.908(18)        \\
       { } &     0.2867 & -0.67(7)          \\
       { } &     0.3084 & -0.58(3)          \\
       { } &     0.4168 & -0.50(8)          \\
       { } &     0.4576 & -0.42(4)          \\
       { } &     0.6169 & -0.256(14)        \\
\hline
    0.1324 &     0.1480 & -0.928(17)        \\
       { } &     0.1488 & -1.01(11)         \\
       { } &     0.2852 & -0.62(6)          \\
       { } &     0.3084 & -0.53(4)          \\
       { } &     0.4137 & -0.55(7)          \\
       { } &     0.4573 & -0.40(4)          \\
       { } &     0.5350 & -0.21(5)          \\
       { } &     0.6169 & -0.264(15)        \\
\hline
    0.1333 &     0.1463 & -0.85(2)          \\
       { } &     0.2796 & -0.55(9)          \\
       { } &     0.3084 & -0.45(5)          \\
       { } &     0.4029 & -0.53(10)         \\
       { } &     0.6169 & -0.220(18)        \\
\hline
    0.1338 &     0.1447 & -0.74(3)          \\
       { } &     0.3084 & -0.44(7)          \\
       { } &     0.6169 & -0.18(3)          \\
\hline
    0.1342 &     0.1439 & -0.73(4)          \\
       { } &     0.3084 & -0.36(7)          \\
       { } &     0.6169 & -0.18(3)          \\
\end{tabular}
\end{ruledtabular}
\end{table}

\begin{table}
\caption{Magnetic form factor of the neutron at $\beta = 6.2$.} 
\label{tab:clover62neut}
\begin{ruledtabular}
\begin{tabular}{ddd}
\multicolumn{1}{c}{$\kappa $} & \multicolumn{1}{c}{$a^2 Q^2$} & 
\multicolumn{1}{c}{$G_m$}   \\
\hline
    0.1333 &     0.0665 & -1.186(19)        \\
       { } &     0.0667 & -1.10(8)          \\
       { } &     0.1294 & -0.79(4)          \\
       { } &     0.1371 & -0.80(4)          \\
       { } &     0.1892 & -0.61(7)          \\
       { } &     0.2742 & -0.43(2)          \\
\hline
    0.1339 &     0.0661 & -1.07(2)          \\
       { } &     0.1279 & -0.83(8)          \\
       { } &     0.1371 & -0.71(4)          \\
       { } &     0.2742 & -0.37(2)          \\
\hline
    0.1344 &     0.0655 & -1.07(3)          \\
       { } &     0.1259 & -0.62(6)          \\
       { } &     0.1371 & -0.65(6)          \\
       { } &     0.2742 & -0.34(3)          \\
\hline
    0.1349 &     0.0647 & -0.94(3)          \\
       { } &     0.1233 & -0.58(9)          \\
       { } &     0.1371 & -0.55(6)          \\
       { } &     0.2742 & -0.29(3)          \\
\end{tabular}
\end{ruledtabular}
\end{table}

\begin{table}
\caption{Magnetic form factor of the neutron at $\beta = 6.4$.} 
\label{tab:clover64neut}
\begin{ruledtabular}
\begin{tabular}{ddd}
\multicolumn{1}{c}{$\kappa $} & \multicolumn{1}{c}{$a^2 Q^2$} & 
\multicolumn{1}{c}{$G_m$}   \\
\hline
    0.1338 &     0.0375 & -1.208(19)        \\
       { } &     0.0376 & -1.20(7)          \\
       { } &     0.0730 & -0.75(4)          \\
       { } &     0.0771 & -0.80(3)          \\
       { } &     0.1069 & -0.53(4)          \\
       { } &     0.1394 & -0.39(6)          \\
       { } &     0.1542 & -0.44(2)          \\
\hline
    0.1342 &     0.0373 & -1.17(3)          \\
       { } &     0.0724 & -0.81(9)          \\
       { } &     0.0771 & -0.79(6)          \\
       { } &     0.1542 & -0.35(3)          \\
\hline
    0.1346 &     0.0370 & -1.07(3)          \\
       { } &     0.0713 & -0.58(5)          \\
       { } &     0.0771 & -0.64(4)          \\
       { } &     0.1034 & -0.42(5)          \\
       { } &     0.1338 & -0.28(6)          \\
       { } &     0.1542 & -0.35(2)          \\
\hline
    0.1350 &     0.0366 & -1.00(3)          \\
       { } &     0.0700 & -0.58(8)          \\
       { } &     0.0771 & -0.59(6)          \\
       { } &     0.1009 & -0.44(11)         \\
       { } &     0.1140 & -0.37(6)          \\
       { } &     0.1542 & -0.31(2)          \\
\hline
    0.1353 &     0.0360 & -0.85(6)          \\
       { } &     0.0681 & -0.52(12)         \\
       { } &     0.0771 & -0.55(9)          \\
       { } &     0.1542 & -0.25(4)          \\
\end{tabular}
\end{ruledtabular}
\end{table}

\clearpage

\section{}

The following tables contain the results of our dipole fits. 
The masses are given in lattice units.

\begingroup
\squeezetable
\begin{table*}[h]
\caption{Dipole fits of the isovector form factors.}
\label{tab:clover.dipisov}
\begin{ruledtabular}
\begin{tabular}{ddddddd}
\multicolumn{1}{c}{$\kappa $} & \multicolumn{1}{c}{$aM_e$} & 
\multicolumn{1}{c}{$A_m$} & \multicolumn{1}{c}{$aM_m$} & 
\multicolumn{1}{c}{$aM_1$} & \multicolumn{1}{c}{$A_2$} &  
\multicolumn{1}{c}{$aM_2$}  \\
\hline
\multicolumn{7}{c}{$\beta = 6.0$} \\
\hline
    0.1320 & 0.657(6)          & 4.3(2)           
           & 0.66(2)           & 0.756(7)         
           & 3.4(2)            & 0.61(3)           \\
    0.1324 & 0.637(5)          & 4.5(2)           
           & 0.64(2)           & 0.745(7)         
           & 3.6(2)            & 0.60(3)           \\
    0.1333 & 0.593(8)          & 4.3(3)           
           & 0.61(3)           & 0.705(9)         
           & 3.5(4)            & 0.56(4)           \\
    0.1338 & 0.570(12)         & 4.1(6)           
           & 0.57(5)           & 0.675(14)        
           & 3.4(7)            & 0.50(6)           \\
    0.1342 & 0.534(14)         & 4.0(7)           
           & 0.57(7)           & 0.633(17)        
           & 4.(2)             & 0.44(15)          \\
\hline
\multicolumn{7}{c}{$\beta = 6.2$} \\
\hline
    0.1333 & 0.493(4)          & 4.79(17)         
           & 0.507(16)         & 0.579(5)         
           & 3.87(18)          & 0.480(19)         \\
    0.1339 & 0.477(5)          & 4.6(2)           
           & 0.484(17)         & 0.566(6)         
           & 3.7(2)            & 0.45(2)           \\
    0.1344 & 0.441(6)          & 4.7(3)           
           & 0.46(2)           & 0.539(8)         
           & 3.9(4)            & 0.42(3)           \\
    0.1349 & 0.411(7)          & 4.3(3)           
           & 0.45(3)           & 0.511(9)         
           & 3.5(4)            & 0.41(3)           \\
\hline
\multicolumn{7}{c}{$\beta = 6.4$} \\
\hline
    0.1338 & 0.375(3)          & 5.16(18)         
           & 0.358(10)         & 0.436(4)         
           & 4.19(19)          & 0.340(12)         \\
    0.1342 & 0.358(5)          & 5.2(3)           
           & 0.347(17)         & 0.422(7)         
           & 4.2(3)            & 0.33(2)           \\
    0.1346 & 0.333(4)          & 5.1(3)           
           & 0.322(14)         & 0.399(5)         
           & 4.2(3)            & 0.299(16)         \\
    0.1350 & 0.310(5)          & 4.8(4)           
           & 0.319(16)         & 0.384(7)         
           & 3.9(4)            & 0.30(2)           \\
    0.1353 & 0.282(11)         & 3.7(5)           
           & 0.35(4)           & 0.350(14)        
           & 2.9(9)            & 0.33(9)           
\end{tabular}
\end{ruledtabular}
\end{table*}
\endgroup

\begingroup
\squeezetable
\begin{table*}[b]
\caption{Dipole fits of the proton form factors.}
\label{tab:clover.dipprot}
\begin{ruledtabular}
\begin{tabular}{ddddddd}
\multicolumn{1}{c}{$\kappa $} & \multicolumn{1}{c}{$aM_e$} & 
\multicolumn{1}{c}{$A_m$} & \multicolumn{1}{c}{$aM_m$} & 
\multicolumn{1}{c}{$aM_1$} & \multicolumn{1}{c}{$A_2$} &  
\multicolumn{1}{c}{$aM_2$}  \\
\hline
\multicolumn{7}{c}{$\beta = 6.0$} \\
\hline
    0.1320 & 0.673(5)          & 2.59(12)         
           & 0.66(2)           & 0.720(5)         
           & 1.62(14)          & 0.61(4)           \\
    0.1324 & 0.653(5)          & 2.71(13)         
           & 0.64(2)           & 0.706(5)         
           & 1.70(14)          & 0.61(3)           \\
    0.1333 & 0.608(6)          & 2.59(18)         
           & 0.62(3)           & 0.665(7)         
           & 1.61(19)          & 0.59(5)           \\
    0.1338 & 0.583(10)         & 2.5(3)           
           & 0.58(5)           & 0.635(10)        
           & 1.7(5)            & 0.51(8)           \\
    0.1342 & 0.543(11)         & 2.4(4)           
           & 0.59(7)           & 0.596(11)        
           & 1.5(4)            & 0.55(10)          \\
\hline
\multicolumn{7}{c}{$\beta = 6.2$} \\
\hline
    0.1333 & 0.505(4)          & 2.93(10)         
           & 0.505(16)         & 0.547(4)         
           & 1.97(12)          & 0.48(2)           \\
    0.1339 & 0.485(4)          & 2.81(12)         
           & 0.483(17)         & 0.527(4)         
           & 1.83(13)          & 0.45(3)           \\
    0.1344 & 0.453(5)          & 2.87(19)         
           & 0.46(2)           & 0.501(6)         
           & 2.0(2)            & 0.42(3)           \\
    0.1349 & 0.420(5)          & 2.6(2)           
           & 0.44(3)           & 0.469(6)         
           & 1.7(2)            & 0.41(4)           \\
\hline
\multicolumn{7}{c}{$\beta = 6.4$} \\
\hline
    0.1338 & 0.383(3)          & 3.15(11)         
           & 0.359(10)         & 0.412(3)         
           & 2.13(12)          & 0.340(15)         \\
    0.1342 & 0.366(4)          & 3.17(19)         
           & 0.349(17)         & 0.398(5)         
           & 2.1(2)            & 0.33(2)           \\
    0.1346 & 0.342(3)          & 3.08(18)         
           & 0.325(14)         & 0.374(4)         
           & 2.1(2)            & 0.30(2)           \\
    0.1350 & 0.319(4)          & 3.0(2)           
           & 0.314(16)         & 0.356(4)         
           & 2.0(3)            & 0.29(2)           \\
    0.1353 & 0.287(7)          & 2.2(3)           
           & 0.36(4)           & 0.326(8)         
           & 1.5(4)            & 0.33(5)           
\end{tabular}
\end{ruledtabular}
\end{table*}
\endgroup

\begin{table}
\caption{Dipole fits of the neutron magnetic form factor.}
\label{tab:clover.dipneut}
\begin{ruledtabular}
\begin{tabular}{ddd}
\multicolumn{1}{c}{$\kappa $} & 
\multicolumn{1}{c}{$A_m$} & \multicolumn{1}{c}{$aM_m$}  \\
\hline
\multicolumn{3}{c}{$\beta = 6.0$} \\
\hline
    0.1320 & -1.68(9)          & 0.65(2)           \\
    0.1324 & -1.78(9)          & 0.62(2)           \\
    0.1333 & -1.71(14)         & 0.59(3)           \\
    0.1338 & -1.6(2)           & 0.57(6)           \\
    0.1342 & -1.6(3)           & 0.55(7)           \\
\hline
\multicolumn{3}{c}{$\beta = 6.2$} \\
\hline
    0.1333 & -1.90(8)          & 0.498(18)         \\
    0.1339 & -1.78(9)          & 0.48(2)           \\
    0.1344 & -1.90(15)         & 0.44(3)           \\
    0.1349 & -1.70(16)         & 0.43(3)           \\
\hline
\multicolumn{3}{c}{$\beta = 6.4$} \\
\hline
    0.1338 & -2.05(8)          & 0.351(11)         \\
    0.1342 & -2.06(16)         & 0.34(2)           \\
    0.1346 & -2.01(14)         & 0.314(15)         \\
    0.1350 & -1.84(16)         & 0.319(19)         \\
    0.1353 & -1.5(2)           & 0.32(4)           
\end{tabular}
\end{ruledtabular}
\end{table}

\end{document}